\documentclass[twocolumn,showpacs,preprintnumbers,amsmath,amssymb,aps,prb]
{revtex4}
\usepackage{epsf}
\usepackage{graphicx}
\usepackage{dcolumn}
\usepackage{bm}

\begin{document}

\title{Quasiperiodic functions theory and the superlattice
potentials for a two-dimensional electron gas.}

\author{Andrei Ya. Maltsev}
 \email{maltsev@itp.ac.ru}
\affiliation{ L.D.Landau Institute for Theoretical Physics,
117940 ul. Kosygina 2, Moscow}

\date{\today}

\begin{abstract}

 We consider Novikov problem of the classification
of level curves of quasiperiodic functions on the plane
and its connection with the conductivity of two-dimensional
electron gas in the presence of both orthogonal magnetic field
and the superlattice potentials of special type. We show
that the modulation techniques used in the recent papers on the
2D heterostructures permit to obtain general quasiperiodic
potentials for 2D electron gas and consider the asymptotic limit
of conductivity when $\tau \rightarrow \infty$. Using the theory 
of quasiperiodic functions we introduce here the topological 
characteristics of such potentials observable in the conductivity.
The corresponding characteristics are the direct analog of
the ``topological numbers'' introduced previously in the
conductivity of normal metals.  

\end{abstract}

\pacs{02.40.-k, 05.45.-a, 72.20My}

\maketitle

\section{Introduction.} 

 In the present paper we consider modern experimental techniques 
of potential modulation for the two-dimensional electron gas 
and show that they permit to obtain
quasiperiodic potentials on the plane with different numbers
of quasiperiods. Then we use topological results concerning 
the geometry of the level curves of such potentials 
(S.P. Novikov problem) to obtain the asymptotic 
($\tau \rightarrow \infty$) behavior of the conductivity phenomena
in these systems. Namely, we consider the quasiclassical approach,
where the quasiclassical cyclotron orbits drift along the level
curves of potential in the presence of magnetic field $B$
which makes the geometry of such level curves important for
transport phenomena. Our approach is based on the topological
methods used previously (by S.P. Novikov and the author) in the
theory of normal metals (\cite{novmal1,novmal2,malnov3}) 
and the quasiclassical description of the
transport phenomena in high-mobility 2D electron gas introduced
by C.W.J. Beenakker (\cite{beenakker}) for the explanation of
new oscillations in $B$-dependence of conductivity found in the
periodically modulated 2D electron gas (\cite{WKPW}).

 We will consider here the cases of potentials with 3 and 4 
quasiperiods and use a set of rather deep topological
theorems concerning S.P. Novikov problem obtained during the last
years. Let us say here that these two cases are actually the only 
cases which were studied seriously in topology 
and where very nice results were obtained.
Namely, the full classification of the non-closed level
curves was obtained for the case of potentials with 3 quasiperiods 
on the plane and it was shown (\cite{dynn3,dynn7}) that only
the so-called ``topologically regular'' level curves appear in the 
case when the non-closed level curves exist in a non-zero
energy interval $\epsilon_{1} \leq V({\bf r}) \leq \epsilon_{2}$.
The corresponding curves reveal nice geometrical properties
being bounded by straight strips of a finite width in the
plane and passing through them. Moreover, it can be shown that the
mean directions of these strips always correspond to some topological
numbers characterizing the potential $V({\bf r})$. Thus for the case
of 3 quasiperiods these numbers can be represented as the
indivisible integer triples $(m_{1}, m_{2}, m_{3})$ which can be 
defined experimentally from the mean directions of potential level 
curves. For the case of 4 quasiperiods the corresponding numbers
are $4$-tuples which can again be defined from the mean 
directions of the topologically regular open level curves in the 
transport phenomena. However, in the last case the existence of 
topologically regular open level curves can be stated only for
small perturbations of purely periodic potentials in 
${\mathbb R}^{2}$ (\cite{novikov5}).

 In this paper we show that special modulations of 2D
electron gas give quasiperiodic potentials on the plane and
introduce the corresponding topological numbers and their
connection with the modulation pictures. Let us say that the
topological numbers of this kind were introduced already
in the theory of normal metals
(\cite{novmal1}, \cite{novmal2}, \cite{malnov3}) where the
``geometric strong magnetic field limit'' in the galvanomagnetic
phenomena was considered. For this case only the situation with
3 quasiperiods was important and the topological numbers had 
the form of integer triples $(m_{1}, m_{2}, m_{3})$.
Another feature of the situation in the normal metals is that 
just the Fermi energy level $\epsilon_{F}$ is important for 
the asymptotic behavior of conductivity in the ``geometric limit''.

As we already said, we will use here the ``drifting orbits''
approximation and consider the case $\tau \rightarrow \infty$ which
corresponds to the ``geometric limit'' in the situation of 2D
electron gas. We consider in details the electric
conductivity tensor $\sigma^{ik}$ in the asymptotic form
for $\tau \rightarrow \infty$ when a strong anisotropy of 
$\sigma^{ik}$ reveals the mean directions of topologically
regular trajectories and gives the corresponding topological
numbers. 

 Let us say also that the cases of chaotic behavior of the
potential level curves are also possible for the quasiperiodic
potentials $V({\bf r})$ (\cite{tsarev,dynn4}). 
The asymptotic behavior of $\sigma^{ik}$ is more complicated 
in this case and we will
not consider it here in details. For the case of 3 quasiperiods,
however, the generic behavior of conductivity should correspond 
to topologically regular situation and the chaotic cases are
``exclusive'' unlike the cases with big numbers of quasiperiods.

\section{Basic definitions and historical notes.}

 According to the standard definition a quasiperiodic
function $f({\bf r})$, ${\bf r} \in {\mathbb R}^{n}$
with $N$ quasiperiods $(N \geq n)$ is a restriction of a
periodic function $F({\bf R})$, ${\bf R} \in {\mathbb R}^{N}$
with $N$ linearly independent periods 
${\bf l}_{1}, \dots, {\bf l}_{N}$ in a bigger linear space
${\mathbb R}^{N}$ to some ``plane'' 
${\mathbb R}^{n} \subset {\mathbb R}^{N}$. The corresponding
subspace ${\mathbb R}^{n}$ can then be given by a linear
system

$$ \left\{ \begin{array}{c}
a_{11} y^{1} + a_{12} y^{2} + \dots + a_{1N} y^{N} = b_{1} 
\cr
\dots
\cr
a_{N-n,1} y^{1} + a_{N-n,2} y^{2} + \dots + a_{N-n,N} y^{N} 
= b_{N-n} \end{array} \right. $$

 We will say that the plane ${\mathbb R}^{n}$ has the maximal
irrationality if it is not parallel to any vector ${\bf l}$
belonging to the lattice $L$ generated by vectors 
${\bf l}_{1}, \dots, {\bf l}_{N}$:

$$L = \{ p_{1} {\bf l}_{1} + \dots + p_{N} {\bf l}_{N} \,\,\, , 
\,\,\,\,\, p_{1}, \dots, p_{N} \in {\mathbb Z}\}$$

 We will call the plane 
${\mathbb R}^{n} \subset {\mathbb R}^{N}$ rational if it
contains (i.e. parallel to) exactly $n$ linearly independent 
vectors belonging to $L$.

 Obviously the generic planes ${\mathbb R}^{n}$ in 
${\mathbb R}^{N}$ have the maximal irrationality. It is easy to 
see also that any vector ${\bf l} \in L$ parallel to the
plane ${\mathbb R}^{n}$ in a non-generic situation becomes
a period of the function $f({\bf r})$ in ${\mathbb R}^{n}$.
The function $f({\bf r})$ corresponding to a rational plane
${\mathbb R}^{n} \subset {\mathbb R}^{N}$ is a $n$-periodic 
function in ordinary sense. It is easy to see also that 
generic quasiperiodic function $f({\bf r})$ with $N$ quasiperiods 
has no periods in ${\mathbb R}^{n}$ for $N > n$.

 We are going to consider the case $n = 2$ such that the function
$f({\bf r}) = f(x,y)$ is a quasiperiodic function on the 
two-dimensional plane ${\mathbb R}^{2}$. Namely, we will
describe here important features of the global geometry of 
the level curves 
$f({\bf r}) = const$ (Novikov problem) which will play the 
main role for the phenomena discussed in this paper.

 Let us say here that the Novikov problem is still unsolved
for the case of arbitrary $N > 2$ and we are going to deal here
with the cases $N = 3$ and $N = 4$ where new topological
and physical results were obtained during the last years 
(see \cite{novikov1,zorich,novikov2,novikov3,dynn1,dynn3,
tsarev,zorich2,novikov4,novmal1,dynn5,dynn6,dynn4,dynmal,
malts,novmal2,dynn7,novikov5,novikov6,RdLeo,malnov3}).
According to the definition, the corresponding functions 
$f({\bf r})$ will be the restrictions of periodic functions
in ${\mathbb R}^{3}$ and ${\mathbb R}^{4}$ on some 
two-dimensional planes ${\mathbb R}^{2}$. Let us say here
some words about this situation. 

 We will start with the very
important case $N = 3$ where the full classification of the
curves $f({\bf r}) = const$ is constructed now. This case
plays extremely important role for the galvanomagnetic
phenomena in normal metals 
(see \cite{novmal1,novmal2,malnov3}) where
the function ${\hat f}({\bf p})$, ${\bf p} = (p_{1},p_{2},p_{3})$
is defined in the space of quasimomenta of the Bloch 
electron in the crystal. The function $f({\bf p})$ is a
restriction of the three-periodic function ${\hat f}({\bf p})$
on a two-dimensional plane (orthogonal to the magnetic field)
embedded in ${\mathbb R}^{3}$. The level curves of $f({\bf p})$
are the intersections of the corresponding plane 
$\Pi = {\mathbb R}^{2}$ with the 3-periodic two-dimensional
level surfaces of the (smooth) function ${\hat f}({\bf p})$
(dispersion relation).
From the physical point of view the level curves of the function
$f$ are the quasiclassical electron trajectories in the 
${\bf p}$-space in the presence of magnetic field ${\bf B}$.
We have in this case a one-parametric family of planes $\Pi$
orthogonal to ${\bf B}$ and a one-parametric set of the
quasiperiodic functions defined in the different planes.
The form of trajectory in the coordinate space is defined
in this case by it's form in the ${\bf p}$-space keeping
all the main features of global geometry. For instance
the projection of orbit on $xy$-plane in ${\bf r}$-space
coincides precisely with the trajectory in ${\bf p}$-space
rotated by $\pi/2$. Let us also point out that only the 
trajectories close to the Fermi level are important for the 
case of normal metals.

 The importance of the geometry of these trajectories for the
galvanomagnetic phenomena was pointed out in 
\cite{lifazkag,lifpes1,lifpes2} 
(see also survey articles \cite{lifkag1,lifkag2})
where also the first examples
of concrete two-dimensional periodic Fermi surfaces in 
${\mathbb R}^{3}$ were considered. The problem of full
classification of such trajectories was set by S.P.Novikov in
\cite{novikov1} and considered later in his school
(A.V.Zorich, I.A.Dynnikov, S.P.Tsarev).

 Let us say here that this problem is rather complicated 
already for $N = 3$ and required non-trivial methods
based on topology and dynamical systems theory to be solved
completely. The most important breakthroughs in this problem
were made in (\cite{zorich}) and (\cite{dynn3}) where 
very important topological theorems about the non-closed
trajectories were proved.

 Based on this methods the ``topological quantum characteristics''
observable in the conductivity of normal metals were introduced
in \cite{novmal1}. These characteristics arise from the
geometry of the Fermi surface and have the form of triples
of integer numbers connected with the asymptotic behavior of
conductivity for $B \rightarrow \infty$ (see also the survey
articles \cite{novmal2,malnov3}). For these physical 
phenomena an additional property pointed out in \cite{novmal1}
and called later the ``Topological resonance'' played the important 
role. We will see here how all these properties can be revealed
in the two-dimensional electron gas in a quasiperiodic potential 
$V({\bf r})$.

 Recently the full classification of different trajectories 
in this situation was finished by I.A.Dynnikov 
(\cite{dynn4,dynn7}) which permits to describe the
total picture of the asymptotic behavior of conductivity
for $B \rightarrow \infty$ in normal metals with arbitrary
complicated dispersion relations (\cite{malnov3}).

 The case $n = 2$, $N = 4$ was started by S.P.Novikov in
\cite{novikov5} where a deep topological theorem analogous
to the result of \cite{zorich} for this situation was proved. 
Let us point out 
here that the case $N = 4$ looks very complicated from
topological point of view and this theorem is the only
deep topological result in this case up to now.

 In this paper, however, we work with the coordinate space
rather than with the momenta space and  
consider quasiperiodic functions $V({\bf r})$ 
where ${\bf r} = (x, y)$ plays the role of 
the ordinary coordinate vector on the plane. In this 
situation only one plane ${\mathbb R}^{2}$ embedded in
${\mathbb R}^{3}$ or ${\mathbb R}^{4}$ will be important.
However, also the global characteristics of the total
family of potentials corresponding to different parallel
planes will arise through the action of the ``quasiperiodic
group'' as we will see below.

\section{The quasiclassical trajectories and 2D electron gas.}

 Let us introduce first the notations for different
level curves of potential $V({\bf r})$ according to
\cite{dynn4,novmal2,dynn7,malnov3}.
We will assume now that the function $V({\bf r})$ is a
Morse function on ${\mathbb R}^{2}$, i.e. all the critical
points of $V({\bf r})$ ($\nabla V({\bf r}) = 0$) are
non-degenerate 
($\det ||\partial_{i}\partial_{j} V|| \neq 0$). All the
critical points of $V({\bf r})$ can then be just the 
non-degenerate local minima, the non-degenerate saddle points 
or the non-degenerate local maxima. 
The local geometry of the level curves close
to these critical points are shown at Fig 1, a-c.

\begin{figure}
\epsfxsize=1.0\hsize
\epsfbox{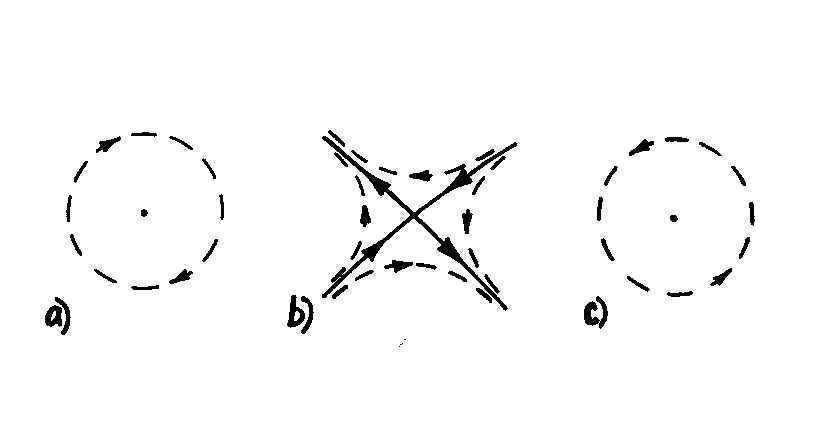}
\caption{\label{fig1} The level curves of the function 
$V({\bf r})$ close to the local minimum, the saddle-point and
the local maximum of $V({\bf r})$.}
\end{figure}

 Let us call now the level curves of $V({\bf r})$ the
quasiclassical drift trajectories according to our 
further considerations.\footnote{Let us use here the
word ``orbit'' for the circular cyclotron electron orbit
in the magnetic field ${\bf B}$ and ``trajectory'' for
the drift of the center of cyclotron orbit in the presence
of $V({\bf r})$. We hope that there should be no 
misunderstanding because of two similar terms.} 
We will also put formally the arrows on the level curves
according to the direction of drift in the magnetic field.

\vspace{0.5cm}

{\bf Definition 1.} {\it We call a trajectory non-singular if it
is not adjacent to a critical (saddle) point of the function
$V({\bf r})$. The trajectories adjacent to critical points
as well as the critical points themselves we call singular
trajectories (see Fig. 1).}

\vspace{0.5cm}

{\bf Definition 2.} {\it We call a non-singular trajectory compact 
if it is closed on the plane. We call a non-singular trajectory
open if it is unbounded in ${\mathbb R}^{2}$.}

\vspace{0.5cm}

 The examples of singular, compact and open non-singular
trajectories are shown on the Fig. 2, a-c.

\begin{figure}
\epsfxsize=1.0\hsize
\epsfbox{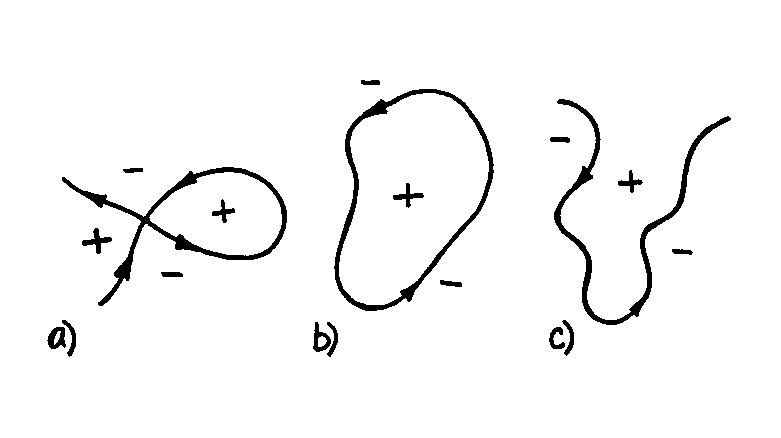}
\caption{\label{fig2} The singular, compact and open non-singular
quasiclassical trajectories. The signs $"+"$ and $"-"$ show the
regions of larger and smaller values of $V({\bf r})$ respectively.}
\end{figure}

 It is easy to see also that singular trajectories have the 
measure zero among all the trajectories on the plane.

 The geometry of compact trajectories will not be interesting
for us here since we are going to consider the ``geometric''
limit corresponding to the long lifetime between two
scattering processes. In this limit we assume that every
center of drifting cyclotron orbit belongs to the same 
trajectory for rather long time. This means in particular
that all compact trajectories will be passed many times before 
jumping to another trajectory due to the scattering act.
This situation corresponds precisely to the ``geometric
strong magnetic field limit'' considered in 
\cite{lifazkag,lifpes1,lifpes2,novmal1,novmal2,malnov3} 
where the conductivity in normal 
metals was studied. However, in our situation this geometric
limit does not correspond to strong magnetic field limit
as we will see below.
 
\vspace{0.5cm} 
 
 {\bf Definition 3.} {\it We call an open trajectory topologically
regular (corresponding to ``topologically integrabl'' case)
if it lies within a straight line of finite width
in ${\mathbb R}^{2}$ and passes through it from $-\infty$ to
$\infty$ (see Fig. 3, a). All other open trajectories we will
call chaotic (Fig. 3, b).}

\vspace{0.5cm}

\begin{figure}
\epsfxsize=1.0\hsize
\epsfbox{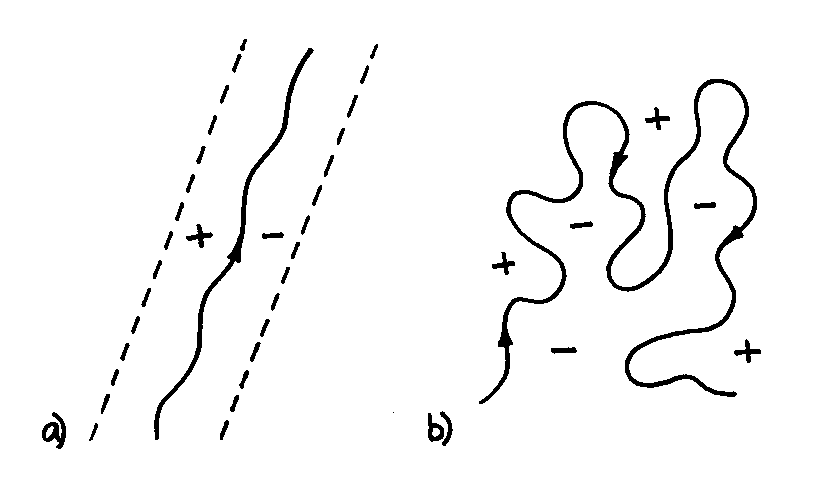}
\caption{\label{fig3} ``Topologically regular'' (a) and ``chaotic'' 
(b) level curves of the function $V({\bf r})$ in the plane
${\mathbb R}^{2}$.}
\end{figure}

 In the simple case of a periodic function $V({\bf r})$ ($N = 2$) 
all open trajectories are periodic and we have only
``topologically regular'' case according to our classification.
However, in the quasiperiodic case the situation is much more
complicated and the chaotic trajectories can exist already for
$N = 3$ (\cite{tsarev,dynn4}). These special trajectories can 
reveal rather complicated stochastic behavior for the general 
quasiperiodic potentials but fortunately  the ``generic'' open 
trajectories are still topologically regular for the case $N = 3$.
Let us point out that this fact was formulated first by 
S.P.Novikov in form of conjecture and 
plays now (together with ``topological resonance'') the crucial 
role for topological phenomena in normal metals 
(\cite{novmal1,novmal2,malnov3}). 
Here we are also going to
consider mainly the ``topologically regular'' situation for
$N = 3$ and $N = 4$ and we will show that the same 
``topological numbers'' can be observed also for two-dimensional
electron gas in specific potentials 
(quasiperiodic superlattices) built by 
special experimental techniques in 2-dimensional structures.

 Let us describe now the quasiclassical approach for the
two-dimensional electron gas which we are going to consider.

 The quasiclassical consideration of the 2D electron gas
in the presence of rather strong magnetic field $B$ and a
potential $V({\bf r})$ was presented in \cite{beenakker}
in connection with the oscillations of conductivity discovered 
in \cite{WKPW}. The experiment in \cite{WKPW} 
(D.Weiss, K.v.Klitzing, K.Ploog, G.Weimann) used the
holographic illumination of high mobility $AlGaAS-GaAS$
heterojunctions at the temperatures $T \leq 4.2 K$. 
The expanded laser beam was splitten into two parts which gave an 
interference picture with a period $a$ on a two-dimensional
sample. The magnetic field ${\bf B}$ was directed normally to
the sample and the electron behavior was determined by the 
magnetic field and the additional periodic potential 

$$V({\bf r}) = V(x) \,\,\,\,\, , \,\,\,\,\,
V(x+a) = V(x)$$
arising after the holographic illumination. The amplitude
of $V(x)$ was much smaller than the Fermi energy of the system.
Measuring the resistivity in the both directions along and
perpendicular to the interference fringes the authors of 
\cite{WKPW} found the magnetoresistance oscillations in
$1/B$ for magnetic fields smaller than needed for 
Shubnikov-de-Haas oscillations.

 This phenomenon was explained by C.W.J. Beenakker in
\cite{beenakker} from the quasiclassical consideration
and called the ``Commensurability oscillations''. According
to the quasiclassical approach the potential $V(x)$ should be
averaged over a quasiclassical electron cyclotron orbit
with radius $r_{B} = mv_{F}/eB$ on the Fermi level to get
the effective averaged potential 
${\bar V}(x,B) = V_{B}^{eff}(x)$ depending on the magnetic
field $B$. The condition of weakness of potential
$V(x)$ ($eV_{rms}/\epsilon_{F} \ll 1$, where
$V_{rms}$ is the root mean square of $V(x)$)
should be imposed in this situation.
The drift of the center of cyclotron orbit is given
then by the equation

\begin{equation}
\label{dynsyst}
{d {\bf r}_{0} \over dt} = {e \over B^{2}}
\left[ \nabla V_{B}^{eff}({\bf r}_{0}) \times {\bf B} \right]
\end{equation}

 According to (\ref{dynsyst}) we have a drift of the centers
of cyclotron orbits along the level curves of 
$V_{B}^{eff}({\bf r}) = V_{B}^{eff}(x)$ with the speed 
proportional to $||\nabla V_{B}^{eff}({\bf r})||$ on these  
curves. As was pointed out in \cite{beenakker},
the drifting motion gives an anisotropic contribution 
to the conductivity in the plane, depending on the potential
$V_{B}^{eff}(x)$. The crucial role for the magnetoresistance
oscillations is played then by a strong dependence of 
$V_{B}^{eff}({\bf r})$ on the value of $B$ connected with the
commensurability of the cyclotron radius $r_{B}$ 
(for a given Fermi energy) and the period of potential $a$.
The corresponding contribution to the conductivity was thus the
oscillating function of $1/B$ due to the periodic 
commensurability $2 r_{B} = ka$ with some integer $k$. 

 The explicit formulae for the conductivity was obtained
in \cite{beenakker} for the model potential having the form
$V(x) = V_{x} cos \, 2\pi x/a$. Obviously the main features
of this picture will also be true for many generic periodic
potentials $V(x)$. Let us also give here the references on the 
papers 
\cite{GerWeiKl,WinKotPl,VasPeet,StrMD,ZhanGer,PeetVas1,WKPW2,
BetDelMai,weiss,GerWeiWul,gerhardts1,PeetVas2,
TBrSim,AlSmWei,KSFvKE,Nog} 
where different questions connected with this problem were
considered (we are sorry for impossibility to give here
the complete list of works on this area).

 Let us consider now the works where the situation of
potentials $V({\bf r})$ modulated both in $x$ and $y$
directions was considered. The potential 
$V({\bf r}) = V(x,y)$ was induced in this case by two
independent sets of interference fringes parallel to 
the $x$ and $y$ axes and the potential $V(x,y)$ was 
a periodic function in ${\mathbb R}^{2}$ with two
periods given by vectors $(a,0)$ and $(0,a)$.

 As was found experimentally 
(\cite{WKPW2,weiss,GerWeiWul}) the
additional modulation in $y$ direction suppresses the
commensurability oscillations in this case. The quasiclassical
consideration of this situation was made in \cite{GrLonDav}
where again the drift of electron orbits along the constant 
energy levels of potential $V(x,y)$ was considered. Two types
of the drift trajectories were considered in \cite{GrLonDav}:

 1) the ``pinned orbits'' (corresponding to compact energy
level curves);

 2) the ``drifting orbits'' (corresponding to unbounded energy
level curves in the plane).

 As was assumed in \cite{GrLonDav} only the contribution of the
``drifting orbits'' was important for the commensurability 
oscillations in this case and the ``pinned orbits'' were
unessential for this phenomenon. According to this assumption
the suppression of the commensurability oscillations can be
explained by the appearance of the ``pinned orbits'' for the
potentials modulated both in $x$ and $y$ directions. 
Unlike the case of potentials
modulated just in $x$ direction, a new condition that the
compact trajectories are passed many times by the centers
of cyclotron orbits between two scattering acts appeared in
\cite{GrLonDav}. This requirement is similar to the
condition of the ``geometric strong magnetic field limit''
considered in \cite{lifazkag,lifpes1,lifpes2} for normal
metals. However, the limit $B \rightarrow \infty$ does
not correspond to the geometric limit in this situation
and only $\tau \rightarrow \infty$ should be considered
as the geometric limit for this case. Easy to see also that
only periodic ``drifting orbits'' can appear for purely
periodic potentials $V(x, y)$.

 Let us also point out here that the analytic dependence 
of the resistance on the value of $B$ was also calculated in
\cite{GrLonDav} in the interesting interval for 
model potentials having few harmonics. This dependence is 
more complicated compared with the case of 1D modulated
potentials but still reveals the effect of commensurability
also in this situation. The mean directions of trajectories
appeared in \cite{GrLonDav} were parallel to $x$ and $y$ axes
and to the diagonal $y = - x$ in different examples.
As was also pointed out in \cite{GrLonDav} the 
``drifting orbits'' can exist only for potentials with
broken rotational symmetry which explains the maximal
suppression of the commensurability oscillations for
the case of equal modulation intensity in both $x$ and $y$
directions.

 In this paper we will not consider in detail the 
$B$-dependence of conductivity for our more complicated 
potentials since it should reveal much more complicated 
behavior in this case. Instead we are going to consider
the geometric properties of conductivity tensor in the
limit $\tau \rightarrow \infty$ arising from the global
geometry of non-singular open trajectories. Namely, we will 
show that this type of potentials can be considered as a
particular case of the quasiperiodic potentials with
fixed number of quasiperiods and use the results
obtained for the Novikov problem to get the 
``topological characteristics'' of the conductivity in
this case. Let us say that this type of ``topological
quantities'' arises in completely different way compared
for example with the Hall effect and characterizes the
geometry of asymptotic of conductivity tensor (but not 
its absolute values).
 
 We will not also put any special conditions on potential
$V(x,y)$ except the quasiperiodic properties. The formulated
results will have a general topological form valid for 
generic potentials $V(x,y)$.

 Before we start the geometric consideration we want to say
also that the holographic illumination is not the unique way
to produce the superlattice potentials for the two-dimensional
electron gas. Let us
mention here the works \cite{AlBetHen,IASLNT,ISMS,FanStil,TIBAS,
PMMLK,WKPW2,GerWeiWul,DavLar,LarDavLonCus,DavPetLon} where
different techniques using the biasing of the specially made 
metallic gates and the piezoelectric effect were considered. Both
1D and 2D modulated potentials as well as more general
periodic potentials with square and hexagonal geometry
appeared in this situation. We want 
to point out that quasiperiodic potentials can be made also
by these techniques in the same way using the superposition of 
several 1D modulations. Actually these techniques give even more
possibilities to produce potentials of different types
even for the quasiperiodic situation. For example a 
superposition of a general periodic potential with a generic 
1D modulation will give quasiperiodic potentials with 3 
quasiperiods which are more general than made just by 3
interference pictures. Also superpositions of two general
periodic potentials on the plane will give a class of the
quasiperiodic potentials with 4 quasiperiods more general
than those which we will consider in detail here. However, we 
would like for simplicity to restrict ourselves to the simpler 
pictures of superpositions of 1D modulation pictures which give 
already all the features of general behavior. We will also use 
everywhere the term ``interference picture'' for the modulation 
pictures. The general geometrical results will then be true for 
the other techniques also.

\section{Novikov problem and the geometric limit for the
case of 3 quasiperiods.}

 Let us come now to Novikov problem and start the topological 
consideration of the level curves of quasiperiodic functions.

 We will first describe the situation for an arbitrary periodic
potential $V(x,y)$ with some periods 
${\bf l}_{1}, {\bf l}_{2} \in {\mathbb R}^{2}$. This picture
is rather simple from the topological point of view but it is
convenient to give it here just to introduce the notations and
to show the general approach which we are going to use. Let us
consider the generic periodic function $f({\bf r})$ on 
${\mathbb R}^{2}$ with the values belonging to some
interval $[f_{min},f_{max}]$. We are interested in the form
of the level curves $f({\bf r}) = c$ where 
$f_{min} \leq c \leq f_{max}$. It is easy to see that for the
values of $c$ close to minimal or maximal value of $f$ all
such level curves are just small closed loops bounding small 
regions of lower or higher values of $f$ (see Fig 4, a,b).

\begin{figure}
\epsfxsize=1.0\hsize
\epsfbox{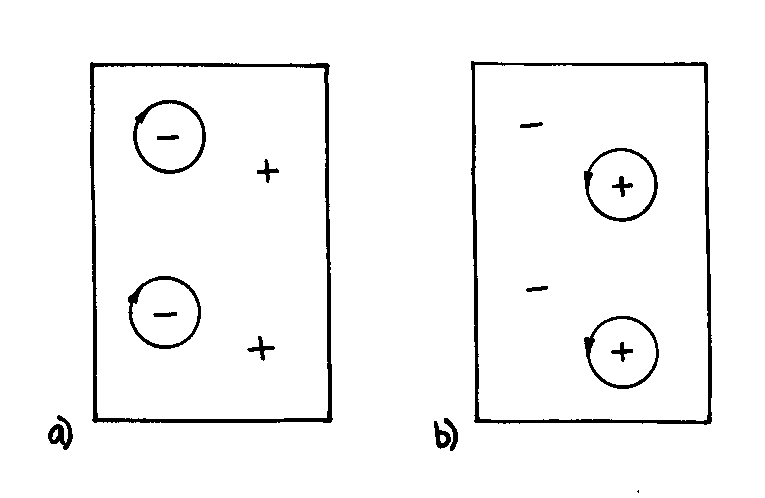}
\caption{\label{fig4} The level curves of $f({\bf r})$ close
to the minimal and maximal values of $f$.}
\end{figure}

 It is not difficult to prove also that the extended trajectories 
(singular or non-singular) always exist in some closed connected
``energy interval'' $f_{1} \leq c \leq f_{2}$
($f_{min} < f_{1} \leq f_{2} < f_{max}$). In generic situation
we have $f_{1} < f_{2}$ but for special functions $f({\bf r})$
also the case $f_{1} = f_{2}$ is possible. This fact is 
actually true for any quasiperiodic function and does not depend
on the number of quasiperiods (the proof in \cite{dynn4}
given for $N = 3$ works actually for any $N$ without any change).
Every non-singular open trajectory is periodic for a periodic 
function $f({\bf r})$ with the mean direction given by some
integral vector ${\bf l} = m_{1} {\bf l}_{1} + m_{2} {\bf l}_{2}$
of lattice generated by periods ${\bf l}_{1}, {\bf l}_{2}$.
We can see then that every non-singular open trajectory for
periodic $f({\bf r})$ corresponds to ``topologically regular''
case.

 It is easy to see also that there can be only a finite number 
of energy levels for a periodic Morse function where the 
singular trajectories can exist. We can claim then that the
non-singular open trajectories always exist in the generic case
$f_{1} < f_{2}$. The opposite statement is also true since
the non-singular open trajectories are stable with respect to
a small change of energy level. The typical situation of
the generic case with the layers of open trajectories is shown
on Fig. 5.

\begin{figure}
\epsfxsize=1.0\hsize
\epsfbox{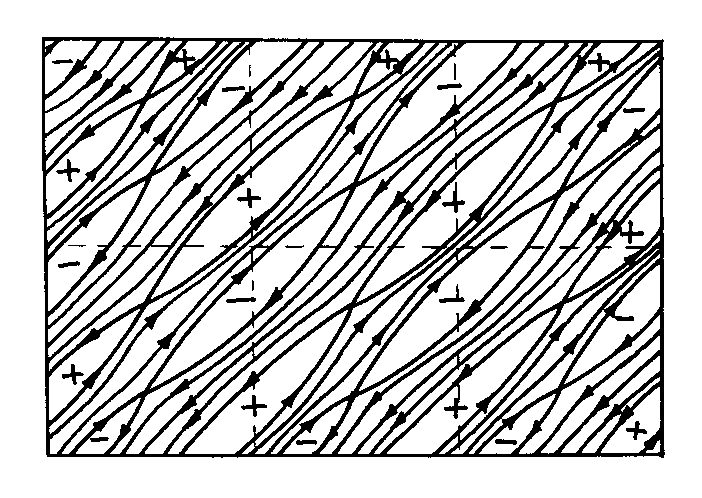}
\caption{\label{fig5} The layers of open periodic trajectories
with the ``non-trivial'' opposite directions ($(1,1)$ and $(-1,-1)$)
for the generic periodic function $f({\bf r})$.}
\end{figure}

 All the open trajectories do not intersect each other and
have a common mean direction passing in both ``direct'' and
the ``opposite'' way.

 The opposite non-generic case $f_{1} = f_{2}$ corresponds to
the absence of the non-singular open trajectories in the plane.
The typical picture for $f_{1} = f_{2}$ is a ``singular net''
on the level $f({\bf r}) = f_{1} = f_{2}$ and the closed
trajectories at all the other levels (Fig. 6). 
Let us pay here a special attention to the last fact to
compare this situation with the more complicated quasiperiodic 
case.

\begin{figure}
\epsfxsize=1.0\hsize
\epsfbox{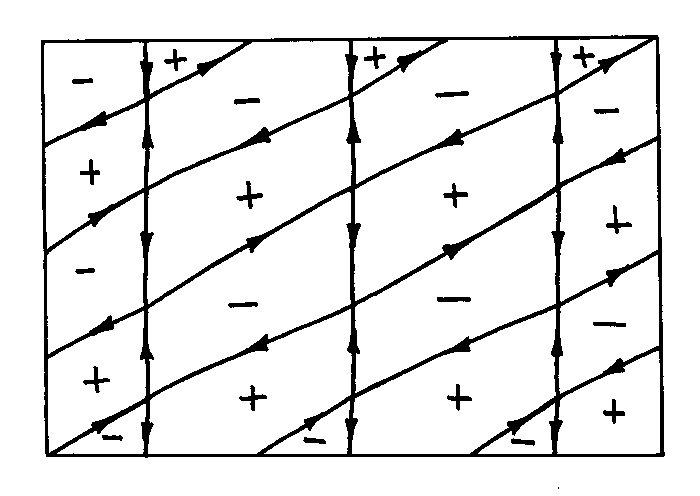}
\caption{\label{fig6} The singular periodic net on the level
$f({\bf r}) = f_{1} = f_{2}$ for a non-generic periodic function
$f({\bf r})$.}
\end{figure}

 It follows also that the case $f_{1} = f_{2}$ always takes 
place for potentials with any kind of rotational symmetry since 
non-singular open trajectories can not exist in this situation.

 Let us give here also the references on the work 
\cite{BrDobPan} where the nice quantization picture
based on the topology of periodic quasiclassical drift 
trajectories in the magnetic field was considered.

 The generic periodic potentials $V({\bf r})$ arise in the 
experiments described above when two independent interference 
pictures with arbitrary directions of interference fringes
are present at the same sample. The potential
$V({\bf r})$ is a functional of the total intensity of
radiation $I({\bf r})$ and has the same periodicity for
any (even nonlocal) translationally invariant dependence
of $V({\bf r})$ on the field $I({\bf r}^{\prime})$. 
For simplicity we will put the requirement that the functional 
$V({\bf r})[I]$ has the variational derivative 
$\delta V({\bf r})/\delta I({\bf r}^{\prime})$ decreasing for
large enough $|{\bf r} - {\bf r}^{\prime}|$. We assume also that
the functional $V({\bf r})[I]$ is smooth, i.e. gives the
smooth function $V({\bf r})$ for any smooth distribution
$I({\bf r}^{\prime})$.

 Let us now come to our main purpose and consider  
potentials $V({\bf r})$ having more complicated form. 
Let us have now three independent interference pictures
on the plane with three different generic directions
of fringes $\bm{\eta}_{1}, \bm{\eta}_{2}, \bm{\eta}_{3}$ 
and periods $a_{1}, a_{2}, a_{3}$ (see Fig. 7).

\begin{figure}
\epsfxsize=1.0\hsize
\epsfbox{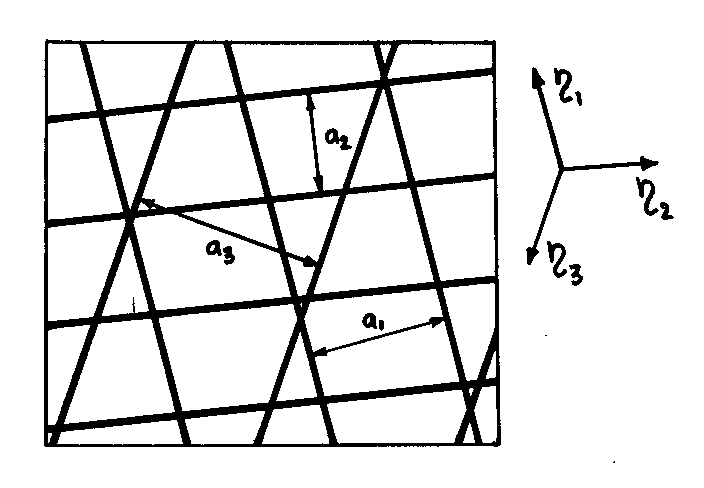}
\caption{\label{fig7} The schematic sketch of three 
independent interference pictures on the plane with different 
periods and intensities.}
\end{figure}

 The total intensity $I({\bf r})$ will be the sum of
intensities 

$$I({\bf r}) = I_{1}({\bf r}) + I_{2}({\bf r}) + 
I_{3}({\bf r})$$
of the independent interference pictures.

 We assume that there are at least two non-coinciding 
directions (say $\bm{\eta}_{1}, \bm{\eta}_{2}$) among the set
$(\bm{\eta}_{1}, \bm{\eta}_{2}, \bm{\eta}_{3})$.

 Let us draw three straight lines $q_{1}$, $q_{2}$, $q_{3}$
with the directions $\bm{\eta}_{1}, \bm{\eta}_{2}, \bm{\eta}_{3}$
and choose the ``positive'' and ``negative'' half-planes for every line
$q_{i}$ on the plane. Let us consider now three linear functions 
$X({\bf r})$, $Y({\bf r})$, $Z({\bf r})$ on the plane which are 
the distances from the point ${\bf r}$ to the lines
$q_{1}$, $q_{2}$, $q_{3}$ with the signs $"+"$ or $"-"$
depending on the half-plane for the corresponding line $q_{i}$
(Fig. 8).

\begin{figure}
\epsfxsize=1.0\hsize
\epsfbox{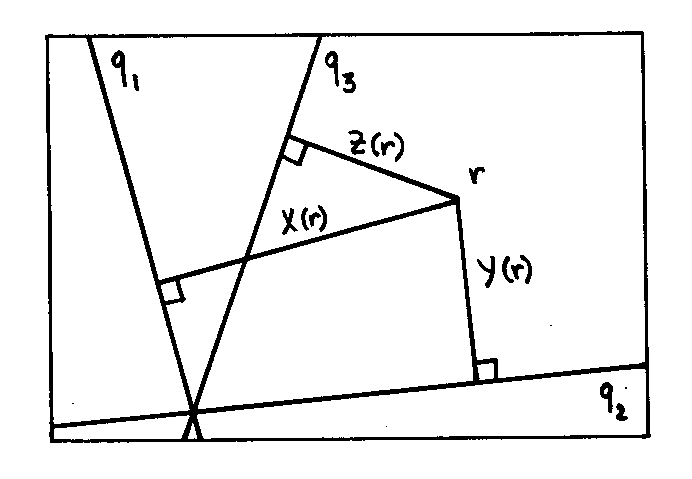}
\caption{\label{fig8} The coordinates $X({\bf r})$, $Y({\bf r})$
and $Z({\bf r})$ on the plane.}
\end{figure}

 The coordinates 

$${\bf R}({\bf r}) = 
\left(X({\bf r}), Y({\bf r}), Z({\bf r}) \right)$$
give now a parametric representation of our plane 
$\Pi^{2} = {\mathbb R}^{2}$ in the 3-dimensional space 
${\mathbb R}^{3}$. The total intensity $I({\bf r})$ can
be considered then as the restriction to $\Pi^{2}$ 
of the periodic function ${\hat I}(X,Y,Z)$:

$${\hat I}(X,Y,Z) = I_{1}(X) + I_{2}(Y) + I_{3}(Z)$$ 
corresponding to the lattice in ${\mathbb R}^{3}$ generated
by vectors $(a_{1}, 0, 0)$, $(0, a_{2}, 0)$,
$(0, 0, a_{3})$.
The plane $\Pi^{2}$ passes through the origin according
to Fig. 8 (although it is not necessary if the lines 
$q_{1}$, $q_{2}$, $q_{3}$ do not intersect at one point
in ${\mathbb R}^{2}$).

 Let us point out here that the standard inner product on
the plane $\Pi^{2}$ does not coincide with the product in
${\mathbb R}^{3}$ in this construction (from the metric
point of view the plane ${\mathbb R}^{2}$ will be linearly
deformed in the embedding ${\bf R} = {\bf R}(x, y)$). However, 
the inner product will not be important at all in our further
considerations so we don't pay any attention to this fact.
Let us just say that it's possible to introduce a special
inner product in ${\mathbb R}^{3}$ such that it's restriction
on $\Pi^{2}$ will give the standard metric in ${\mathbb R}^{2}$.
Nevertheless, all the topological statements will be invariant
under the group of all non-degenerate linear transformations
and we will not need this construction at all.

 We can define now smooth periodic functions
${\hat V}(X,Y,Z)$ and ${\hat V}_{B}^{eff}(X,Y,Z)$ in 
${\mathbb R}^{3}$ such that the functions $V(x,y)$
and $V_{B}^{eff}(x,y)$ will be the restrictions of 
${\hat V}(X,Y,Z)$ and ${\hat V}_{B}^{eff}(X,Y,Z)$
on the plane $\Pi^{2}$. Indeed, consider any point
${\bf R} = (X,Y,Z) \in {\mathbb R}^{3}$. Let us draw
a two-dimensional plane $\Pi^{2\prime}$ through the
point ${\bf R}$ parallel to the plane $\Pi^{2}$. We have
then the total intensity $I^{\prime}({\bf R})$ in the plane
$\Pi^{2\prime}$ defined as the restriction of 
${\hat I}(X,Y,Z)$ on $\Pi^{2\prime}$. Let us define now the 
functions ${\hat V}(X,Y,Z)$ and ${\hat V}_{B}^{eff}(X,Y,Z)$
at the point ${\bf R}$ as the corresponding functions defined 
in the plane $\Pi^{2\prime}$ passing through 
${\bf R}$ using the functional $V({\bf r})[I]$ and the
averaging over the cyclotron orbits in $\Pi^{2\prime}$.
Easy to see that the functions 
${\hat V}(X,Y,Z)$, ${\hat V}_{B}^{eff}(X,Y,Z)$ are 
smooth periodic functions in ${\mathbb R}^{3}$ with
periods $(a_{1}, 0, 0)$, $(0, a_{2}, 0)$, $(0, 0, a_{3})$.
Obviously the functions ${\hat V}|_{\Pi^{2}}$ and
${\hat V}_{B}^{eff}|_{\Pi^{2}}$ give the required potential
$V({\bf r})$ and the effective potential $V_{B}^{eff}({\bf r})$
in the initial two-dimensional plane ${\mathbb R}^{2}$.

 Let us introduce now an important definition of the 
``quasiperiodic group'' acting on the potentials described
above. As we saw, our construction gives us an embedding 
$\Pi^{2}$ of the initial plane ${\mathbb R}^{2}$ in the
three-dimensional space ${\mathbb R}^{3}$. At the same time
we get the additional planes $\Pi^{2\prime}$ in ${\mathbb R}^{3}$
parallel to $\Pi^{2}$ with different $I^{\prime}({\bf r})$,
$V^{\prime}({\bf r})$, $V_{B}^{eff\prime}({\bf r})$ corresponding
to the same ${\hat I}({\bf R})$, ${\hat V}({\bf R})$ and
${\hat V}_{B}^{eff}({\bf R})$. It is easy to see that the functions
$I^{\prime}({\bf r})$, $V^{\prime}({\bf r})$, 
$V_{B}^{eff\prime}({\bf r})$ correspond to the case of 
three interference pictures with the same mean directions of 
fringes and periods
$(\bm{\eta}_{1}, a_{1})$, $(\bm{\eta}_{2}, a_{2})$,
$(\bm{\eta}_{3}, a_{3})$ but with shifted positions of maxima 
and minima for every interference picture.

\vspace{0.5cm}

 {\bf Definition 4.} {\it We will say that all the potentials
$V^{\prime}({\bf r})$ (as well as $V_{B}^{eff\prime}({\bf r})$
for every given $B$) are related by a ``quasiperiodic group''
of transformations.}

\vspace{0.5cm}

 According to the Definition 4 we define the action of
a ``quasiperiodic group'' in ${\mathbb R}^{2}$ as the parallel
shifts of the plane $\Pi^{2}$ in the space ${\mathbb R}^{3}$.
The ``quasiperiodic group'' is then a 3-parametric Abelian group
isomorphic to 3-dimensional torus 
${\mathbb T}^{3} = {\mathbb R}^{3}/L$

$$L = m_{1} (a_{1}, 0, 0) + m_{2} (0, a_{2}, 0) +
m_{3} (0, 0, a_{3}) $$

$$(m_{1}, m_{2}, m_{3}) \in {\mathbb Z}^{3} $$ 
containing the (non-compact) algebraic subgroup of ordinary
translations in ${\mathbb R}^{2}$.

 As we will see below, this definition will be very convenient
in consideration of open trajectories for potentials of
this type in ${\mathbb R}^{2}$. Namely, we will see that all
the global properties of open trajectories will be the same
for all potentials related by the ``quasiperiodic group''
in the case of generic $(\bm{\eta}_{1}, a_{1})$, 
$(\bm{\eta}_{2}, a_{2})$, $(\bm{\eta}_{3}, a_{3})$.
In other words, for generic $(\bm{\eta}_{1}, a_{1})$,
$(\bm{\eta}_{2}, a_{2})$, $(\bm{\eta}_{3}, a_{3})$
the global geometry of open trajectories will not depend
on the positions of minima and maxima of the interference
pictures and will be defined just by the set
$(\bm{\eta}_{1}, a_{1})$, $(\bm{\eta}_{2}, a_{2})$, 
$(\bm{\eta}_{3}, a_{3})$ and the intensities 
$I_{1}$, $I_{2}$, $I_{3}$ (although the potentials 
$V({\bf r})$, $V_{B}^{eff}({\bf r})$ will
be different in these cases). Let us say, however, that this
property can be broken
for special $(\bm{\eta}_{1}, a_{1})$, 
$(\bm{\eta}_{2}, a_{2})$, $(\bm{\eta}_{3}, a_{3})$
corresponding to purely rational directions of $\Pi^{2}$ in
${\mathbb R}^{3}$.

 According to the previous definition we will say that a
quasiperiodic potential has irrationality 3 or maximal
irrationality if it has no periods in ${\mathbb R}^{2}$. 
We will say that the potential $V({\bf r})$ has irrationality 
2 if it has only one (up to the integer multiplier)
period in ${\mathbb R}^{2}$. We will say that the potential
$V({\bf r})$ has irrationality 1 if it has two linearly
independent periods in ${\mathbb R}^{2}$.\footnote{We are
sorry for a little bit unnatural definitions of irrationality.
We follow here the standard definitions arising from $3D$
topology approach.} As can be easily seen the last case
corresponds to the purely periodic potentials $V({\bf r})$. 
It is easy to see also that the potentials $V({\bf r})$ of
irrationality 3, 2 and 1 correspond to the cases when
the plane $\Pi^{2}$ contains no vectors belonging to $L$,
just one (up to the integer multiplier) vector belonging to $L$ 
and two linearly independent vectors belonging to $L$ respectively.
Obviously all the potentials related by the ``quasiperiodic
group'' have the same irrationality in the plane.

 Let us discuss now briefly the connection of irrationality
with the directions and periods of interference pictures in our 
situation.

 We assume as previously that there are at least two
different directions of the interference fringes in our picture.
The picture given by two corresponding sets of interference
fringes is purely periodic in ${\mathbb R}^{2}$ with the periods
${\bf u}_{1}$, ${\bf u}_{2}$ parallel to
$\bm{\eta}_{1}$ and $\bm{\eta}_{2}$ respectively
(see Fig. 9).

\begin{figure}
\epsfxsize=1.0\hsize
\epsfbox{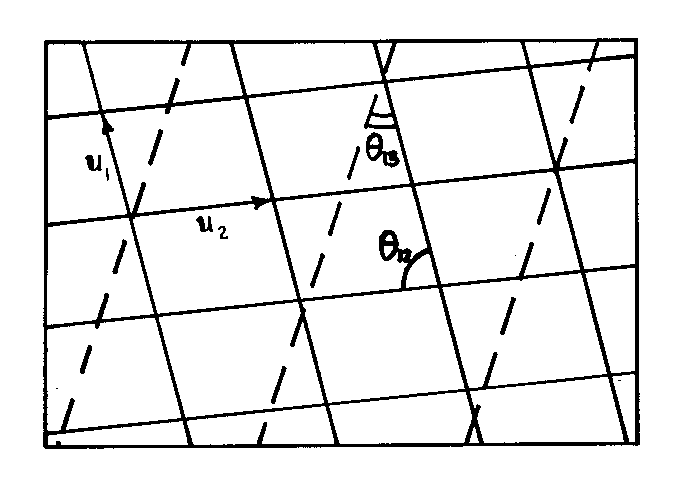}
\caption{\label{fig9} The periodic picture formed by two sets
of parallel interference fringes with common directions
$\bm{\eta}_{1}$, $\bm{\eta}_{2}$ and the added third set with
direction $\bm{\eta}_{3}$.}
\end{figure}

  We can see then that the total picture has a period in 
${\mathbb R}^{2}$ if some nontrivial integer linear combination
$$m_{1} {\bf u}_{1} + m_{2} {\bf u}_{2}\,\,\, , 
\,\,\,\,\, (m_{1},m_{2}) \in {\mathbb Z}^{2}/(0,0)$$
of periods ${\bf u}_{1}$, ${\bf u}_{2}$
leaves invariant also the third interference picture 
corresponding to pair $(\bm{\eta}_{3}, a_{3})$.

 The corresponding condition for 
$m_{1} {\bf u}_{1} + m_{2} {\bf u}_{2}$ can then
be written in the form

\begin{equation}
\label{taucond}
(m_{1} {\bf u}_{1} + m_{2} {\bf u}_{2},
\bm{\xi}_{3}) = k a_{3} \,\,\, ,
\,\,\,\,\, k \in {\mathbb Z}
\end{equation}
where $\bm{\xi}_{3}$ is a unit vector orthogonal to
$\bm{\eta}_{3}$ in the plane.

 The equation (\ref{taucond}) has no nontrivial solutions
in generic situation and can be satisfied only for special
$\bm{\eta}_{3}$ and $a_{3}$. It's not difficult to show
that for purely rational potentials $V({\bf r})$ (two linearly
independent solutions of (\ref{taucond})) the direction 
$\bm{\eta}_{3}$ should also correspond to an integer 
vector in the lattice $L^{\prime}$ generated by vectors 
${\bf u}_{1}$, ${\bf u}_{2}$:

$$L^{\prime} = \{
m_{1} {\bf u}_{1} + m_{2} {\bf u}_{2} \,\,\, ,
\,\,\,\,\, (m_{1}, m_{2}) \in {\mathbb Z}^{2} \} $$

 We can put for this case 
$\bm{\eta}_{3} \sim
m_{1} {\bf u}_{1} + m_{2} {\bf u}_{2}$
for some integer $m_{1}$, $m_{2}$.
Also the corresponding period $a_{3}$ should satisfy to 
a special condition in this situation.
If we introduce the angles $\theta_{12}$, $\theta_{13}$
between the directions 
$\bm{\eta}_{1}$, $\bm{\eta}_{2}$ and
$\bm{\eta}_{1}$, $\bm{\eta}_{3}$,
$0 < \theta_{12} \leq \pi/2$, $0 < \theta_{13} \leq \pi$
(Fig. 9) we can get the relations for $\theta_{13}$ and 
$a_{3}$ which define all the pairs 
$(\bm{\eta}_{3}, a_{3})$ corresponding to purely
rational potentials $V({\bf r})$:

\begin{equation}
\label{etacond}
tg \, \theta_{13} = {m_{2} a_{1} sin \, \theta_{12} \over
m_{1} a_{2} - m_{2} a_{1} cos \, \theta_{12}}
\end{equation}

\begin{equation}
\label{acond}
k_{3} a_{3} \, sin \, \theta_{12} = k_{1} a_{1} \, sin (\theta_{12} 
+ \theta_{13}) + k_{2} a_{2} \, sin \, \theta_{13}
\end{equation}
where $m_{1}, m_{2}, k_{1}, k_{2}, k_{3} \in {\mathbb Z}$,
$(m_{1}, m_{2}) \neq (0,0)$, 
$(k_{1}, k_{2}) \neq (0,0)$, $k_{3} \neq 0$.

 For the case of just one period (irrationality 2) we can
have either the condition (\ref{etacond}) for $\theta_{13}$
(``rational'' direction of $\bm{\eta}_{3}$) but with
$a_{3}$ not satisfying to (\ref{acond}) or the condition
(\ref{acond}) for $a_{3}$ but with $\theta_{13}$ not 
satisfying to (\ref{etacond}).

 Easy to see that both cases of irrationality 1 and 2
have the measure zero among all potentials constructed by
three arbitrary interference pictures.

 As we already said above the case of irrationality 1
corresponds to purely periodic potentials $V(x,y)$. Let us
however make here some remark. Namely, for arbitrary periodic
potential $V(x,y)$ the corresponding periods ${\bf l}_{1}$,
${\bf l}_{2}$ can be much bigger than the values of $a_{1}$,
$a_{2}$, $a_{3}$. We can conclude then that even the
``topologically regular'' periodic open trajectories can have
rather nontrivial structure on rather long distances since
the period of trajectory is very big. The width of the straight
line containing a periodic trajectory can be also compatible
with periods of $V(x,y)$ in this case being quite big with 
respect to the periods of modulations $a_{1}$, $a_{2}$, $a_{3}$.
Also the rational mean direction of the periodic trajectories can
have ``rather big denominator'' such that this rationality will
not play an essential role in the real picture. Instead, the
typical features observable in generic situation of
irrationality 3 will appear on the distances smaller than the
periods $|{\bf l}_{1}|$, $|{\bf l}_{2}|$ of potential $V(x,y)$.
According to this remark we can actually try to consider 
potentials of irrationality 1 or 2 as generic potentials
of irrationality 3 if the periods of these potentials are rather
big. The special features connected with rationality 
can then be revealed only for 
very big values of $\tau$ such that the free motion length
is much larger than $|{\bf l}_{1}|$, $|{\bf l}_{2}|$.

 Let us formulate now (in our language) the first 
theorem about the open trajectories for the quasiperiodic
potentials $V({\bf r})$ and $V_{B}^{eff}({\bf r})$
with 3 quasiperiods corresponding to the first theorem on 
the Novikov problem proved in \cite{zorich}:

\vspace{0.5cm}

 {\bf Theorem 1.} {\it Consider a purely periodic potential
$V^{(0)}({\bf r})$ (or $V_{B}^{(0)eff}({\bf r})$) 
generated by three independent interference
pictures with some parameters 
$(\bm{\eta}^{(0)}_{1}, a^{(0)}_{1})$,
$(\bm{\eta}^{(0)}_{2}, a^{(0)}_{2})$, 
$(\bm{\eta}^{(0)}_{3}, a^{(0)}_{3})$
satisfying to (\ref{etacond}), (\ref{acond}). Then for all
the potentials $V({\bf r})$ (and $V_{B}^{eff}({\bf r})$)
with parameters 
$(\bm{\eta}_{1}, a_{1})$, $(\bm{\eta}_{2}, a_{2})$,
$(\bm{\eta}_{3}, a_{3})$ close enough to
$(\bm{\eta}^{(0)}_{1}, a^{(0)}_{1})$,
$(\bm{\eta}^{(0)}_{2}, a^{(0)}_{2})$, 
$(\bm{\eta}^{(0)}_{3}, a^{(0)}_{3})$
all the open non-singular electron trajectories will
correspond to topologically regular case only. }

\vspace{0.5cm}

 Using the same methods as in \cite{zorich} it is possible
to prove also that Theorem 1 will be true also for small 
variations of the intensities $I_{1}({\bf r})$, $I_{2}({\bf r})$,
$I_{3}({\bf r})$ of the laser beams and the form of the functional
$V({\bf r})[I]$.

 Let us say here that Theorem 1 makes rather strong statement
about generic potentials close to periodic ones. However,
the corresponding ``stability zones'' for parameters
$(\bm{\eta}_{1}, \bm{\eta}_{2}, \bm{\eta}_{3},
a_{1}, a_{2}, a_{3})$ (and $I_{1}$, $I_{2}$, $I_{3}$)
depend on the initial values of
$(\bm{\eta}^{(0)}_{1}, a^{(0)}_{1})$,
$(\bm{\eta}^{(0)}_{2}, a^{(0)}_{2})$,
$(\bm{\eta}^{(0)}_{3}, a^{(0)}_{3})$
and become very small for large values of periods 
$|{\bf l}_{1}|$, $|{\bf l}_{2}|$ of the initial potential.
Due to this reason Theorem 1 can not say anything about arbitrary
potential $V({\bf r})$ (with 3 quasiperiods) since we can have the 
situation when it does not belong to any stability zone 
corresponding to any rational potential $V^{(0)}({\bf r})$.
Nevertheless, this theorem is very important and we will see also
that only the result of this type can be formulated for more
complicated case of potentials with 4 quasiperiods.

 Let us discuss now the general situation of arbitrary potentials
$V({\bf r})$ with 3 quasiperiods. We will start first with the
generic situation of potentials of irrationality 3 and then
discuss additional features which can arise in the cases of 
irrationality 1 and 2. Let us make here the reference on the 
survey article \cite{dynn7} where the final theorems in
the most complete form were formulated. The development of this 
problem and the considerations of physical phenomena
can be found in 
\cite{novikov1,zorich,novikov2,novikov3,dynn1,dynn3,tsarev,
zorich2,novikov4,novmal1,dynn5,dynn6,dynn4,dynmal,malts,novmal2,
dynn7,novikov5,novikov6,RdLeo,malnov3}. 
Let us say also that all the results in 
\cite{dynn7} and in all previous papers were formulated in another
language using the 3-dimensional topology terminology.
We will not discuss here the topological questions in detail
and just claim that the following statements can be derived from 
the topological theorems formulated in \cite{dynn7}.

\vspace{0.5cm}

 {\bf Theorem 2.} {\it Let us fix the value of $B$ and consider 
the generic quasiperiodic
potential $V_{B}^{eff}({\bf r})$ (of irrationality 3) 
taking the values in some interval 
$\epsilon_{min}(B) \leq V_{B}^{eff}({\bf r}) 
\leq \epsilon_{max}(B)$.
Then:

 1) Open quasiclassical trajectories $V_{B}^{eff}({\bf r}) = c$ 
always exist either in a connected energy interval

$$\epsilon_{1}(B) \leq c \leq \epsilon_{2}(B)$$ 
$(\epsilon_{min}(B) < \epsilon_{1}(B) < 
\epsilon_{2}(B) < \epsilon_{max}(B))$
or just at one energy value $c = \epsilon_{0}(B)$ 
(i.e. $\epsilon_{1}(B) = \epsilon_{2}(B) = \epsilon_{0}(B)$).

 2) For the case of a finite interval 
($\epsilon_{1}(B) < \epsilon_{2}(B)$) all the non-singular open
trajectories correspond to topologically regular case, i.e.
lie in straight strips of a finite width (Fig. 3, a)
and pass through them. All the strips have the same mean directions
for all the energy levels 
$c \in [\epsilon_{1}(B), \epsilon_{2}(B)]$
such that all the open trajectories are in average parallel to
each other for all values of $c$.

 3) The values $\epsilon_{1}(B)$, $\epsilon_{2}(B)$ or 
$\epsilon_{0}(B)$
are the same for all the potentials of irrationality 3
connected by the ``quasiperiodic group''.

 4) For the case of a finite energy interval 
($\epsilon_{1}(B) < \epsilon_{2}(B)$) all the non-singular open 
trajectories also have the same mean direction for all the
potentials (of irrationality 3) connected by the 
``quasiperiodic group''. }

\vspace{0.5cm}

 We can see from Theorem 2 that the ``topologically integrable'' 
situation is typical also for the case
of quasiperiodic functions with 3 quasiperiods
being connected with the generic case 
$\epsilon_{1}(B) < \epsilon_{2}(B)$. Let us say also
that for the case of just one energy level 
($\epsilon_{1}(B) = \epsilon_{2}(B) = \epsilon_{0}(B)$) containing 
open trajectories both the topologically regular and ``chaotic''
behavior of open trajectories are possible (see \cite{dynn4}).
This situation can be compared with the situation of purely
periodic potentials where non-singular periodic open
trajectories always appear in the case of finite energy interval
($\epsilon_{1}(B) > \epsilon_{2}(B)$) but only the periodic 
``singular nets'' are possible for the case 
$\epsilon_{1}(B) = \epsilon_{2}(B) = \epsilon_{0}(B)$. As we see 
here quasiperiodic potentials give another possibility in
the last case. 

 Let us consider now the asymptotic behavior of conductivity
in the case of topologically regular open trajectories when 
$\tau \rightarrow \infty$. According to previous papers
(\cite{beenakker,gerhardts1,GrLonDav}) 
we will divide here the conductivity
tensor in two parts $\sigma^{ik}_{0}(B)$ and
$\Delta \sigma^{ik}(B)$ corresponding to the conductivity
without any potential $V({\bf r})$ and an additional
contribution due to potential  $V({\bf r})$. We have then:

$$\sigma^{ik}_{0}(B) = \sigma^{ik}_{0}(B) +
\Delta \sigma^{ik}(B) $$

 In the approximation of the drifting cyclotron orbits
the parts $\sigma^{ik}_{0}(B)$ and $\Delta \sigma^{ik}(B)$
can be interpreted as caused respectively by the 
(infinitesimally small) difference in the electron
distribution function on the same cyclotron orbit (weak
angular dependence) and the (infinitesimally small) difference 
in the occupation of different trajectories by the centers of
cyclotron orbits at different points of 
${\mathbb R}^{2}$ (on the same energy level) as the linear 
response to the (infinitesimally) 
small external field ${\bf E}$. The
asymptotic $\tau \rightarrow \infty$ of both parts
$\sigma^{ik}_{0}(B)$ and $\Delta \sigma^{ik}(B)$ can then
be written from the same arguments used in
\cite{lifazkag,lifpes1,lifpes2} with some additional
remarks specific for this situation. We will just say here that 
the first part $\sigma^{ik}_{0}(B)$ has the standard asymptotic 
form:

$$\sigma^{ik}_{0}(B) \sim {n e^{2} \tau \over m^{eff}}
\left( \begin{array}{cc}
(\omega_{B} \tau)^{-2} & (\omega_{B} \tau)^{-1} \cr
(\omega_{B} \tau)^{-1} & (\omega_{B} \tau)^{-2}
\end{array} \right) $$
for $\omega_{B} \tau \gg 1$ due to the weak angular dependence
$( \sim 1/\omega_{B} \tau )$ of the distribution function on the
same cyclotron orbit. We have then that the corresponding
longitudinal conductivity decreases for 
$\tau \rightarrow \infty$ in all the directions in 
${\mathbb R}^{2}$ and the corresponding condition is
just $\omega_{B} \tau \gg 1$ in this case.

 For the part $\Delta \sigma^{ik}(B)$ the limit 
$\tau \rightarrow \infty$ should, however, be considered
as the condition that every trajectory is passed for rather
long time by the drifting cyclotron orbits to reveal its global
geometry. Thus another parameter $\tau/\tau_{0}$ where
$\tau_{0}$ is the characteristic time of completion of close
trajectories should be used in this case and we should put 
the condition $\tau/\tau_{0} \gg 1$ to have the asymptotic 
regime for $\Delta \sigma^{ik}(B)$.
In this situation the difference between the
open and closed trajectories  plays the main role and the 
asymptotic behavior of conductivity can be calculated in
the form analogous to that used in \cite{lifazkag,lifpes1,lifpes2} 
for the case of normal metals. Namely:

$$ \Delta \sigma^{ik}(B) \sim 
{n e^{2} \tau \over m^{eff}}
\left( \begin{array}{cc}
({\tau_{0}/\tau})^{2} & \tau_{0}/\tau \cr
\tau_{0}/\tau & ({\tau_{0}/\tau})^{2}
\end{array} \right) $$
in the case of closed trajectories and

$$ \Delta \sigma^{ik}(B) \sim
{n e^{2} \tau \over m^{eff}}
\left( \begin{array}{cc}
* & \tau_{0}/\tau \cr
\tau_{0}/\tau & ({\tau_{0}/\tau})^{2}          
\end{array} \right) $$
($* \sim 1$) for the case of open topologically regular 
trajectories if the $x$-axis coincides with the mean direction 
of trajectories.

 We can see then that only the contribution of open orbits to
$\Delta \sigma^{ik}(B)$ remains in (longitudinal)
conductivity for $\tau \rightarrow \infty$. Let us say that these
formulae give just an asymptotic form of conductivity for
$\tau \rightarrow \infty$. In the more precise form they should
include also the multipliers proportional to the parts of the phase
volume filled by the closed and open trajectories and
an appropriate definition of $m^{eff}$ in this situation. 
We will not, however, consider this part in detail since we will
need only the anisotropy of the tensor $\sigma^{ik}$ in the
``geometric limit''.

 The condition $\tau/\tau_{0} \gg 1$ is much stronger
then $\omega_{B} \tau \gg 1$ in the situation described above
just according to the definition of the slow drift of the cyclotron
orbits. We can keep then just this condition in our further
considerations and assume that the main part of conductivity is
given by $\Delta \sigma^{ik}(B)$ in this limit. Easy to see
also that the magnetic field $B$ should not be ``very strong''
in this case.

 According to the remarks above we can write now the main part
of the conductivity tensor $\sigma^{ik}(B)$ in the limit
$\tau \rightarrow \infty$ for the case of topologically regular 
open orbits. Let us take the $x$ axis along the mean direction
of open orbits and take the $y$-axis orthogonal to $x$. The
asymptotic form of $\sigma^{ik}$, $i, k = 1, 2$ can then
be written as:

\begin{equation}
\label{sigmaik}
\sigma^{ik} \sim 
{n e^{2} \tau \over m^{eff}}
\left( \begin{array}{cc}
* & \tau_{0}/\tau \cr
\tau_{0}/\tau & ({\tau_{0}/\tau})^{2}
\end{array} \right) \,\,\, , \,\,\, 
\tau_{0}/\tau \rightarrow 0 
\end{equation}
where $*$ is some value of order of 1 (constant as
$\tau_{0}/\tau \rightarrow 0$).

 The asymptotic form of $\sigma^{ik}$ makes possible the 
experimental observation of the mean direction of topologically
regular open trajectories if the value $\tau/\tau_{0}$ is rather
big.

 Let us introduce here the ``topological numbers'' characterizing
the regular open trajectories introduced first in \cite{novmal1}
for the case of normal metals. We will give a
topological definition of these numbers using the action of the
``quasiperiodic group'' on the quasiperiodic potentials. Let us
assume for simplicity that the potential $V_{B}^{eff}({\bf r})$
is generic and has irrationality 3. We assume that we have the 
``topologically integrable'' situation where the topologically
regular open trajectories exist in some finite energy interval
$\epsilon_{1}(B) \leq c \leq \epsilon_{2}(B)$. According to 
Theorem 2 the values $\epsilon_{1}(B)$, $\epsilon_{2}(B)$ and 
the mean directions
of open trajectories are the same for all the potentials 
constructed from our potential with the aid of the 
``quasiperiodic group''. It follows also from the topological
picture that all the topologically regular trajectories are
absolutely stable under the action of the ``quasiperiodic group''
(for the case of irrationality 3) and can
just ``crawl'' in the plane for a continuous action of
such transformations.

 Let us make now the following transformation:

 We take the first interference picture 
($(\bm{\eta}_{1}, a_{1})$) and shift 
continuously the interference fringes in the direction 
of $grad \, X({\bf r})$ (orthogonal 
to $\bm{\eta}_{1}$) to the distance $a_{1}$ keeping two other
interference pictures unchanged. Easy to see that we will have 
at the end the same potentials $V(x,y)$ and $V_{B}^{eff}(x,y)$
due to the periodicity
of the first interference picture with period $a_{1}$.
Let us fix now some energy level 
$c \in (\epsilon_{1}(B), \epsilon_{2}(B))$ and look at the evolution
of non-singular open trajectories (for $V_{B}^{eff}(x,y)$)
while making our transformation.
We know that we should have parallel open trajectories in
the plane at every time and the initial picture should coincide
with the final according to the construction. The form of
trajectories can change during the process but their mean
direction will be the same according to Theorem 2.\footnote{
The last property was first mentioned in \cite{novmal1} in the
case of normal metals and called later the ``Topological Resonance''.
This property plays an important role for the conductivity of normal 
metals making possible the experimental observation of ``Topological 
Numbers'' for this case.}

 We can claim then that every open trajectory will be ``shifted''
to another open trajectory of the same picture by our continuous
transformation. It's not difficult to prove that all the 
trajectories will then be shifted by the same number of positions 
$n_{1}$ (positive or negative) which depends on the potential 
$V_{B}^{eff}(x,y)$ (Fig. 10).

\begin{figure}
\epsfxsize=1.0\hsize
\epsfbox{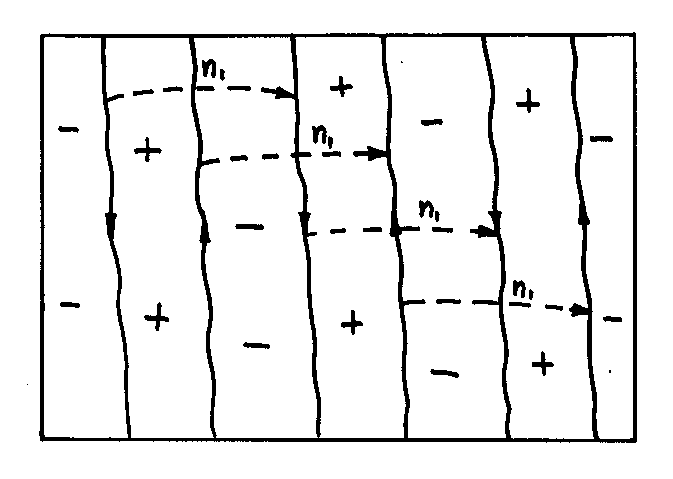}
\caption{\label{fig10} The shift of ``topologically regular''
trajectories by a continuous transformation generated by the
special path in the ``quasiperiodic group''.}
\end{figure}

 The number $n_{1}$ is always even since all the trajectories
appear by pairs with the opposite drift directions.

 Let us now do the same with the second and the third sets of
the interference fringes and get an integer triple
$(n_{1}, n_{2}, n_{3})$ which is a topological characteristic
of potential $V_{B}^{eff}(x,y)$ (the ``positive'' direction of
the numeration of trajectories should be the same for all these
transformations).

 The triple $(n_{1}, n_{2}, n_{3})$ (defined up to the common
sign) can be represented as:

$$(n_{1}, n_{2}, n_{3}) \, = \, M \, (m_{1}, m_{2}, m_{3}) $$
where $M \in {\mathbb Z}$ and $(m_{1}, m_{2}, m_{3})$
is an indivisible integer triple. Both $M$ and
$(m_{1}, m_{2}, m_{3})$ have the topological meaning 
connected with the 
number of connected components carrying open trajectories in
${\mathbb R}^{3}$ and the homological class of every component
in ${\mathbb T}^{3} = {\mathbb R}^{3}/L$ up to the sign.

 Let us mention that for periodic potentials $V(x,y)$ made
just by two interference pictures with common directions 
$\bm{\eta}_{1}$, $\bm{\eta}_{2}$ the corresponding
transformations are actually equivalent
to the shifts along the periods ${\bf u}_{2}$ and ${\bf u}_{1}$
respectively (Fig. 9). It is not difficult to see that the
corresponding numbers $(m_{1}, m_{2})$ are equal then 
(up to the common sign) to $(-i_{1}, i_{2})$ where
$(i_{1}, i_{2})$ is the indivisible integer mean direction
of periodic open trajectories in the lattice $L^{\prime}$
generated by vectors $\{{\bf u}_{1}, {\bf u}_{2}\}$. Easy
to see also that the vectors 
$\{ grad \, X({\bf r})/a_{1}, grad \, Y({\bf r})/a_{2}\}$
give the dual basis to the basis 
$\{{\bf u}_{2}, {\bf u}_{1}\}$ and the mean direction of open 
orbits can be defined from the linear equation

$$m_{1} X({\bf r})/a_{1} + m_{2} Y({\bf r})/a_{2} = 0$$
on the plane.

 It can be proved that the similar situation also takes place
for the topologically regular open trajectories in the case
of quasiperiodic potentials $V({\bf r})$. Let us omit 
here the detailed consideration of the topological picture 
and just say that the common direction of open trajectories
in ${\mathbb R}^{2}$ is defined completely 
by the triple $(m_{1}, m_{2}, m_{3})$. Let us formulate here 
the corresponding statement:

\vspace{0.5cm}

 {\bf Theorem 3.} {\it Consider the functions

$$X^{\prime} ({\bf r}) = X({\bf r})/a_{1} \,\,\, , \,\,\,
Y^{\prime} ({\bf r}) = Y({\bf r})/a_{2} \,\,\, , \,\,\,
Z^{\prime} ({\bf r}) = Z({\bf r})/a_{3} $$
in ${\mathbb R}^{2}$. The mean direction of the regular open
trajectories is given by the linear equation:

\begin{equation}
\label{dircond}
m_{1} X^{\prime}(x,y) + m_{2} Y^{\prime}(x,y) +
m_{3} Z^{\prime}(x,y) = 0
\end{equation}
where $(m_{1}, m_{2}, m_{3})$ is the indivisible integer triple
introduced above. }

\vspace{0.5cm}

 The triples $(m_{1}, m_{2}, m_{3})$ coincide precisely with
the ``Topological Quantum Numbers'' introduced in \cite{novmal1}
for the conductivity in normal metals. Let us say that the
condition (\ref{dircond}) determines completely the numbers
$(m_{1}, m_{2}, m_{3})$ (from the mean direction of open
trajectories) for potentials of irrationality 3. 
This fact permits to extract the values of 
$(m_{1}, m_{2}, m_{3})$ from the direct conductivity
observations using the anisotropy of tensor 
$\sigma^{ik}(B)$. (The formula
(\ref{dircond}) is also true for the case of so-called
``stable'' open trajectories for potentials of irrationality
1 and 2 (see below). The triple 
$(m_{1}, m_{2}, m_{3})$ generally speaking may not be defined 
uniquely from the mean directions of open trajectories in these 
cases and the arguments based on quasiperiodic group play then the
main role in the definition of $(m_{1}, m_{2}, m_{3})$. 
However, it can be measured from the direct conductivity
observations also in these cases due to the stability of these
numbers with respect to the small change of parameters
$(\bm{\eta}_{1}, a_{1})$, $(\bm{\eta}_{2}, a_{2})$,
$(\bm{\eta}_{3}, a_{3})$.)

 A very important property of the integer triples 
$(m_{1}, m_{2}, m_{3})$ is their stability with respect to the
small variations of all the parameters
$\bm{\eta}_{1}$, $\bm{\eta}_{2}$, $\bm{\eta}_{3}$,
$a_{1}$, $a_{2}$, $a_{3}$, $I_{1}$, $I_{2}$, $I_{3}$ and even
of the form of dependence $V({\bf r})[I]$. This means that the
space of parameters 
$(\bm{\eta}_{1}, \bm{\eta}_{2}, \bm{\eta}_{3},
a_{1}, a_{2}, a_{3}, I_{1}, I_{2}, I_{3})$ where the situation
$\epsilon_{2}(B) > \epsilon_{1}(B)$ 
for the energy interval containing
the open trajectories takes place can be divided into different
``stability zones'' $\Gamma_{\alpha}$ where the relations 
(\ref{dircond}) are valid for generic $V_{B}^{eff}({\bf r})$ 
with the same values of 
$(m^{\alpha}_{1}, m^{\alpha}_{2}, m^{\alpha}_{3})$. Let us
emphasize here that the mean directions of open trajectories
are different for different values of parameters even within
the same stability zone $\Gamma_{\alpha}$ and the equation
(\ref{dircond}) gives a fixed relation of these directions
with the directions and periods of the interference fringes
for a given stability zone.

 The zones $\Gamma_{\alpha}$ form an everywhere dense set in the
total space of parameters and in general we can have an infinite 
number of zones parameterized by the numbers
$(m^{\alpha}_{1}, m^{\alpha}_{2}, m^{\alpha}_{3})$. The
triples $(m^{\alpha}_{1}, m^{\alpha}_{2}, m^{\alpha}_{3})$ form
some subset of all possible integer triples $(m_{1}, m_{2}, m_{3})$
(defined up to the common sign)
and give an important topological characteristic of the potentials 
$V_{B}^{eff}({\bf r})$ made by 3 interference pictures. 
The sizes of zones
$\Gamma_{\alpha}$ decrease for the big numbers 
$(m^{\alpha}_{1}, m^{\alpha}_{2}, m^{\alpha}_{3})$ and the
total set $\{\cup \Gamma_{\alpha}\}$ gives a rather complicated 
subset in the space of parameters
$(\bm{\eta}_{1}, \bm{\eta}_{2} , \bm{\eta}_{3},
a_{1}, a_{2}, a_{3}, I_{1}, I_{2}, I_{3})$. Let us say also that
the topologically regular open trajectories are also stable with
respect to any variation of potential $V_{B}^{eff}({\bf r})$ small 
enough which makes possible to observe them also for slightly 
imperfect quasiperiodic potentials $V({\bf r})$.

 Before starting with special possibilities for the periodic
(irrationality 1) or ``partly periodic'' (irrationality 2)
potentials we will say here some words about the ``chaotic'' 
behavior of open trajectories possible in the case
$\epsilon_{1}(B) = \epsilon_{2}(B) = \epsilon_{0}(B)$. Let us say 
that for $\epsilon_{1}(B) = \epsilon_{2}(B)$ both the 
situations of topologically regular and chaotic behavior of open 
trajectories are possible in the quasiperiodic case. The first 
situation always takes place when the corresponding set 
$(\bm{\eta}_{1}, \bm{\eta}_{2} , \bm{\eta}_{3},
a_{1}, a_{2}, a_{3}, I_{1}, I_{2}, I_{3})$ belongs to a
boundary of some stability zone $\Gamma_{\alpha}$ in the
space of parameters. In this case all the non-singular open 
trajectories are topologically regular and correspond to the same
numbers $(m^{\alpha}_{1}, m^{\alpha}_{2}, m^{\alpha}_{3})$.
Another situation arises when the set
$(\bm{\eta}_{1}, \bm{\eta}_{2}, \bm{\eta}_{3},
a_{1}, a_{2}, a_{3}, I_{1}, I_{2}, I_{3})$ is an accumulation 
point for the zones $\Gamma_{\alpha}$ but does not belong to the 
boundary of any $\Gamma_{\alpha}$. In this situation much more
complicated chaotic behavior of open orbits appear at the energy
level $V_{B}^{eff}({\bf r}) = \epsilon_{0}(B)$. Obviously the 
``chaotic'' behavior can be possible only for 
potentials of irrationality 2
or 3. Let us say also that the cases of irrationality 2
(Tsarev chaotic behavior) and 3 (Dynnikov chaotic behavior)
demonstrate completely different types of chaotic behavior in
this situation.

 The first example of chaotic open trajectory was constructed by
S.P.Tsarev (\cite{tsarev,dynn4}) for the case of irrationality 2. 
The corresponding chaotic trajectory, however, has an asymptotic 
direction but can not be bounded by any straight
strip of the finite width in ${\mathbb R}^{2}$. As was later
proved by I.A.Dynnikov (\cite{dynn4}) this situation always takes
place for chaotic trajectories in the case of irrationality 2.
The asymptotic behavior of conductivity tensor reveals also the
strong anisotropy for large $\tau$ in this
situation with slightly different from (\ref{sigmaik}) dependence
on $\tau$ for $\tau \rightarrow \infty$.

 More complicated chaotic trajectories were constructed by 
I.A.Dynnikov (\cite{dynn4}) for the case of irrationality 3
(the approximate form of such kind of trajectories is shown on 
Fig. 3, b). The trajectories of this second kind don't have any
asymptotic direction in ${\mathbb R}^{2}$ ``walking everywhere''
in the plane. The form of conductivity tensor for this type of 
trajectories was suggested in \cite{malts} and is more
complicated then (\ref{sigmaik}). We will not discuss here all the 
details and just say that the conductivity decreases in this
case in all directions for $\tau \rightarrow \infty$ as some
non-integer powers of $\tau$.\footnote{The consideration in
\cite{malts} is based on the purely geometrical aspects and does not 
include the quantum corrections due to the jumps from one part of 
trajectory to another. It can be shown, however, that for chaotic 
trajectories of Dynnikov type these jumps should be also important
for rather big values of $\tau$.} Let us also add here that all the
chaotic trajectories are completely unstable with respect to  
small variations of parameters 
$(\bm{\eta}_{1}, \bm{\eta}_{2}, \bm{\eta}_{3},
a_{1}, a_{2}, a_{3}, I_{1}, I_{2}, I_{3})$ (but remain chaotic
with the ``same geometric properties'' under the action of the
``quasiperiodic group'').

 Let us discuss also the $B$-dependence of tensor
$\sigma^{ik}(B)$ for the limit $\tau \rightarrow \infty$.
The value of $B$ belongs here to some interval
$B_{1} \leq B \leq B_{2}$ such that both the drifting
orbits approximation and the condition
$\tau/\tau_{0} \gg 1$ (as well the absence of quantum
oscillations) are true. The effective potential
$V_{B}^{eff}({\bf r})$ is a function of $B$ in this case and
the geometry of trajectories depends on $B$ through the
potential $V_{B}^{eff}({\bf r})$. Let us just say here that it
can be also proved using topological considerations that the
topologically regular open orbits are also ``locally stable''
with respect to small variations of $B$. However, for rather
big changes of value of $B$ it's possible to have ``jumps''
in this picture and get different mean directions of open
trajectories (as well as chaotic cases) in different
parts of the interval $[B_{1}, B_{2}]$. Let us add also that
the structure of $B$-dependence can be rather complicated
in this case containing an infinite number of small
subintervals with very big numbers $(m_{1},m_{2},m_{3})$
as well as the chaotic cases.

 Actually, all the theorems 1-3 can be reformulated in the same 
form if we add the parameter $B$ to parameters $\bm{\eta}_{1}$,
$\bm{\eta}_{2}$, $\bm{\eta}_{3}$, $a_{1}$, $a_{2}$,
$a_{3}$, $I_{1}$, $I_{2}$, $I_{3}$ introduced above.
The probability of ``jumps'' will then increase for small
stability zones $\Gamma_{\alpha}$ corresponding to big numbers
$(m_{1}, m_{2}, m_{3})$ and the $B$-dependence of $\sigma^{ik}(B)$
will depend strongly on the part of the phase space. The chaotic
trajectories will be completely unstable with respect to any
small variations of $B$. 

 In the same way we can consider the ``stability zones'' 
$\Gamma^{ext}_{\alpha}$ in the extended space of parameters 
including the value of the magnetic field $B$. The total set
$\{\cup \Gamma^{ext}_{\alpha}\}$ will have then an analogous
structure containing in general an infinite number of zones
$\Gamma^{ext}_{\alpha}$ and triples $(m_{1},m_{2},m_{3})$
being everywhere dense in the total set of parameters.

 Let us formulate now the general conjecture of S.P.Novikov
about the chaotic cases for potentials with 3 quasiperiods.
In our situation we will assume that potentials are parameterized
by parameters 

$$(\bm{\eta}_{1}, \bm{\eta}_{2}, \bm{\eta}_{3},
a_{1}, a_{2}, a_{3}, I_{1}, I_{2}, I_{3})$$ 
or 

$$(\bm{\eta}_{1}, \bm{\eta}_{2}, \bm{\eta}_{3},
a_{1}, a_{2}, a_{3}, I_{1}, I_{2}, I_{3}, B)$$
and maybe some
additional parameters characterizing the functional 
$V({\bf r})[I]$.

\vspace{0.5cm}

{\bf Novikov conjecture.} {\it The set of parameters corresponding
to the chaotic behavior of open orbits has measure zero in the
total space of parameters.}

\vspace{0.5cm}

 Let us point out that Novikov conjecture was strictly proved in
the important case when only the quasiclassical trajectories 
belonging to some fixed energy level are taken into account
(\cite{dynn4}). This is precisely the situation
arising in the conductivity in normal metals where only the 
trajectories close to the Fermi surface are important.
The more general situation was also investigated numerically
(\cite{RdLeo}) for the case of the special analytic dispersion
relations where the Novikov conjecture was also confirmed. 
However, the general proof of Novikov conjecture for arbitrary
set of parameters is still unknown.

 We will point out now some additional possibilities which can
arise in the non-generic case of potentials of irrationality
1 or 2 (see \cite{dynn7} for detailed mathematical 
considerations).

 Let us start with the case of irrationality 2 when only one
period ${\bf l}$ (up to an integer multiplier) exists in
${\mathbb R}^{2}$. All the parts (1)-(4) of Theorem 2 are also true
for potentials of irrationality 2. We need, however, to make 
one remark about the situation when the mean direction of the
``topologically regular'' open trajectories coincides with the
period ${\bf l}$ of potential. Easy to see that the open 
trajectories are actually periodic in ${\mathbb R}^{2}$ in this
case with the same period ${\bf l}$. In this 
situation some ``additional pairs'' of periodic open trajectories
can arise and disappear under the action of the ``quasiperiodic
group''. These pairs arise from periodic sets of closed
trajectories under the changing of positions of interference
fringes (with the same 
$(\bm{\eta}_{1}, \bm{\eta}_{2}, \bm{\eta}_{3},  
a_{1}, a_{2}, a_{3})$) and disappear in the same way (Fig. 11).

\begin{figure}
\epsfxsize=1.0\hsize
\epsfbox{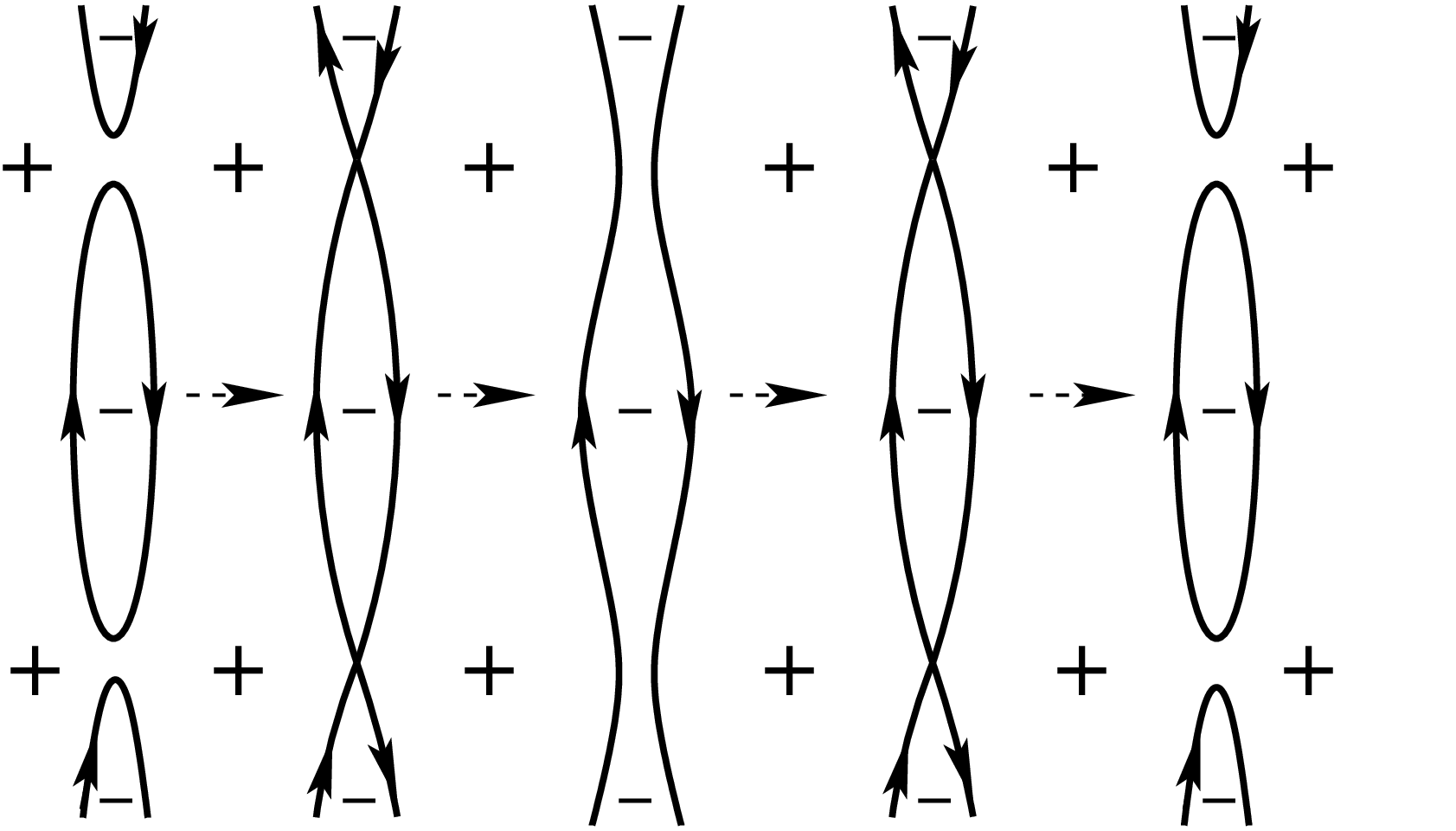}
\caption{\label{fig11} The arising and the disappearance of
periodic trajectories under the action of the 
``quasiperiodic group'' for potentials of irrationality 2 or 1.}
\end{figure}

 The trajectories of this kind are unstable with respect to 
small variations of parameters 
$(\bm{\eta}_{1}, \bm{\eta}_{2}, \bm{\eta}_{3},
a_{1}, a_{2}, a_{3})$ and will be destroyed after any small
variation which does not conserve the period ${\bf l}$
of potential. These trajectories always present
for all the potentials of irrationality 2 connected by the 
``quasiperiodic group'' (on the same energy levels) if they
exist at least for one of them. However, these trajectories can 
``jump'' over the two-dimensional plane ${\mathbb R}^{2}$
disappearing in one place and arising in another under the
action of group transformations. We can call these trajectories
``partly stable'' (or also ``jumping'') in contrary to the absolutely
stable (``crawling'') trajectories described above. It can be proved 
also that the phase volume corresponding to both stable and 
``jumping'' open trajectories is also the same for potentials 
connected by the quasiperiodic group in this situation. 

 The triple of the integer numbers $(n_{1}, n_{2}, n_{3})$ 
can be defined here in the same way as in the 
case of irrationality 3 but these additional pairs of trajectories
should be completely ignored when the action of the 
``quasiperiodic group'' is considered. The motion of stable open
orbits (which always exist in this situation) gives then the
same topological numbers $M$ and $(m_{1}, m_{2}, m_{3})$ as for
close generic potentials.

 All the trajectories still have the same mean direction in this 
situation and the asymptotic form (\ref{sigmaik}) for 
$\tau \rightarrow \infty$ is also true in this case. The formula
(\ref{dircond}) is also valid for the directions of open 
trajectories with the same $(m_{1}, m_{2}, m_{3})$. At the end we
mention that the situation described above can arise only if
the mean directions of stable open orbits coincide with the
period ${\bf l}$ of potential $V_{B}^{eff}({\bf r})$ 
and is absent if it is 
not so. As we also mentioned already the chaotic behavior is also
possible for potentials of irrationality 2 but it is always
simpler than for the irrationality 3 potentials.

 Let us now say some words about the purely periodic potentials
(irrationality 1) which can also appear for special
$(\bm{\eta}_{1}, \bm{\eta}_{2}, \bm{\eta}_{3},   
a_{1}, a_{2}, a_{3})$. As we already said all the open trajectories
are purely periodic in this case and only ``topologically integrable''
situation is possible. We also mentioned already that the extended
trajectories can exist here either in the continuous energy interval
$\epsilon_{1}(B) \leq c \leq \epsilon_{2}(B)$ or just at one energy
level $c = \epsilon_{0}(B)$ (periodic singular nets). All the values
$\epsilon_{1}(B)$, $\epsilon_{2}(B)$, $\epsilon_{0}(B)$, however, 
are not necessarily invariant here with respect to the ``quasiperiodic
group'' action and can be different for different 
potentials connected by the ``quasiperiodic group'' transformations.
Also the mean directions of open orbits can be different for two
potentials belonging to the same orbit of the ``quasiperiodic group''.

 We have then that unlike the cases of irrationality 3 or 2
the positions of interference minima and maxima can be
important here for the conductivity behavior and the parameters 
$(\bm{\eta}_{1}, a_{1})$, $(\bm{\eta}_{2}, a_{2})$, 
$(\bm{\eta}_{3}, a_{3})$ do not determine the picture completely.
It can be proved, however, that a change of the mean directions of
open orbits is possible only
if the case of ``periodic singular net'' takes place at least for one
(actually at least for two) of potentials belonging to the same orbit
of the ``quasiperiodic group''. We can assume then that this situation
takes place only if the periodic potential is prepared specially
to have this property and it does not take place for potentials
with rather big periods ${\bf l}_{1}$, ${\bf l}_{2}$ appeared
``by chance'' in the modulation picture. Thus we can assume that 
periodic potentials with rather big ${\bf l}_{1}$, ${\bf l}_{2}$
arising ``by chance'' can be considered actually as generic 
potentials on the physical level of strictness and do not give any 
special features.

\section{Novikov problem for the case of potentials with
4 quasiperiods.}

 Let us consider now more complicated case when $N = 4$ and we
have a potential made by 4 independent interference pictures
(Fig. 12).

\begin{figure}
\epsfxsize=1.0\hsize
\epsfbox{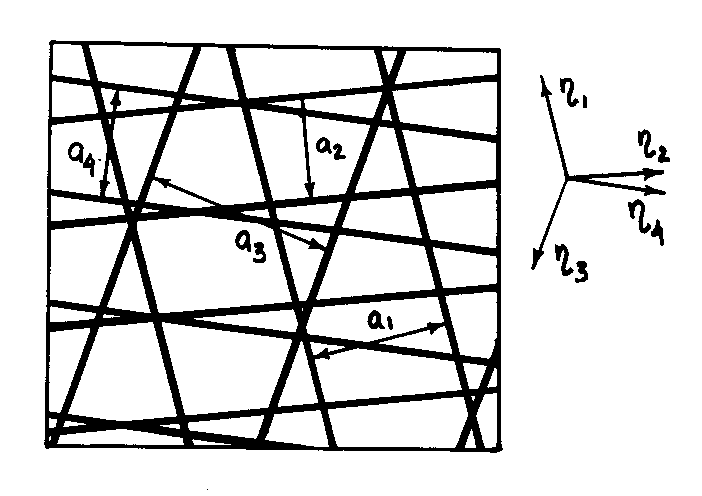}
\caption{\label{fig12} A potential with 4 quasiperiods made 
by 4 independent sets of interference fringes with directions
$\bm{\eta}_{1}$, $\bm{\eta}_{2}$, $\bm{\eta}_{3}$,
$\bm{\eta}_{4}$ and periods $a_{1}$, $a_{2}$, $a_{3}$,
$a_{4}$.}
\end{figure}

 The situation in this case is more complicated than in the case
$N = 3$ and no general classification of open trajectories exists 
at the time. We will present here the theorem of S.P. Novikov
\cite{novikov5} which gives a statement analogous to Zorich
theorem (Theorem 1) in this situation.

 Like in the previous case we define here an embedding of
the plane ${\mathbb R}^{2}$ in the four-dimensional space
${\mathbb R}^{4}$ using the functions
$X({\bf r})$, $Y({\bf r})$, $Z({\bf r})$, $W({\bf r})$
defined in the same way for four interference pictures. We will
need also the functions $X^{\prime}({\bf r})$, 
$Y^{\prime}({\bf r})$, $Z^{\prime}({\bf r})$, $W^{\prime}({\bf r})$
defined as

$$X^{\prime}({\bf r}) = X({\bf r})/a_{1} \,\, , \,\,
Y^{\prime}({\bf r}) = Y({\bf r})/a_{2} \,\, , $$
$$Z^{\prime}({\bf r}) = Z({\bf r})/a_{3} \,\, , \,\,
W^{\prime}({\bf r}) = W({\bf r})/a_{4} $$
(in the same way as previously for the case $N = 3$). 

 The ``total intensity function'' ${\hat I}({\bf R})$, 
${\bf R} \in {\mathbb R}^{4}$ is defined here as

$${\hat I}({\bf R}) = I_{1}(X) + I_{2}(Y) + I_{3}(Z) + I_{4}(W)$$
and is a periodic function with periods $(a_{1},0,0,0)$,
$(0,a_{2},0,0)$, $(0,0,a_{3},0)$, $(0,0,0,a_{4})$ in 
${\mathbb R}^{4}$. The ``big potentials'' ${\hat V}({\bf R})$ 
and ${\hat V}_{B}^{eff}({\bf R})$ are
also defined for every point ${\bf R} \in {\mathbb R}^{4}$ 
through the functional $V({\bf R})[I]$ and the averaging 
over the cyclotron orbits in the plane 
$\Pi^{2\prime} \in {\mathbb R}^{4}$ passing through the point
${\bf R}$ and parallel to the initial plane $\Pi^{2}$. Easy to
see again that the functions ${\hat V}({\bf R})$,
${\hat V}_{B}^{eff}({\bf R})$ are smooth 
4-periodic functions in ${\mathbb R}^{4}$ and the potentials 
$V({\bf r})$, $V_{B}^{eff}({\bf r})$ are the restrictions of 
${\hat V}({\bf R})$ and ${\hat V}_{B}^{eff}({\bf R})$ on the plane
$\Pi^{2}$ embedded in ${\mathbb R}^{4}$. We can define again the 
action of the ``quasiperiodic group'' on the potentials 
$V({\bf r})$, $V_{B}^{eff}({\bf r})$ which is now isomorphic to 
the four-dimensional
torus ${\mathbb T}^{4} = {\mathbb R}^{4}/L$ where
$L$ is an integer lattice generated by vectors $(a_{1},0,0,0)$,
$(0,a_{2},0,0)$, $(0,0,a_{3},0)$, $(0,0,0,a_{4})$.
Let us mention also that the action of this group can be defined 
here in the same way as the shifts of positions of minima and 
maxima of the interference fringes keeping the same the directions
$\bm{\eta}_{1}$, $\bm{\eta}_{2}$, $\bm{\eta}_{3}$, 
$\bm{\eta}_{4}$ and periods 
$a_{1}$, $a_{2}$, $a_{3}$, $a_{4}$.

 Again the statement that the open trajectories always exist
either on the connected energy interval 
$\epsilon_{1}(B) \leq c \leq \epsilon_{2}(B)$ or just at one 
energy level $\epsilon_{0}(B)$ for any 
$V_{B}^{eff}({\bf r})$ is true for the 
case of 4 quasiperiods. It can be also proved that the values 
of $\epsilon_{1}(B)$, $\epsilon_{2}(B)$ or $\epsilon_{0}(B)$ are
the same for generic potentials belonging to the same orbit
of the ``quasiperiodic group''. Moreover, the global behavior
of open trajectories is also the same in this case for all
such potentials and the asymptotic behavior of conductivity
(which is apriori unknown here for the general case)
does not depend on the positions of maxima and minima for the
fixed generic $(\bm{\eta}_{1}, a_{1})$, 
$(\bm{\eta}_{2}, a_{2})$, $(\bm{\eta}_{3}, a_{3})$,
$(\bm{\eta}_{4}, a_{4})$. This properties, however, can
be destroyed for the specially made periodic potentials 
$V({\bf r})$ like in the case of 3 quasiperiods.

 Let us consider now a purely periodic potential $V({\bf r})$ 
formed now by four interference pictures.
We assume again that at least two (say $\bm{\eta}_{1}$,
$\bm{\eta}_{2}$) directions of interference fringes are not   
parallel to each other and give a double-periodic picture in the
plane like in the case of potentials with 3 quasiperiods. Let us
introduce the angles ($\theta_{12}$, $\theta_{13}$, $\theta_{14}$)
between the directions $\bm{\eta}_{1}$ and   
$\bm{\eta}_{2}$, $\bm{\eta}_{3}$, $\bm{\eta}_{4}$
in the same way as in the case of three interference pictures.
From the requirement of periodicity we then will have the same
requirements (\ref{etacond})-(\ref{acond}) for the angles
$\theta_{13}$, $\theta_{14}$ and the periods $a_{3}$, $a_{4}$
with some integer numbers $m^{\prime}_{1}$, $m^{\prime}_{2}$,
$k^{\prime}_{1}$, $k^{\prime}_{2}$, $k^{\prime}_{3}$ 
(for $\theta_{13}$ and $a_{3}$) and $m^{\prime\prime}_{1}$,
$m^{\prime\prime}_{2}$, $k^{\prime\prime}_{1}$, 
$k^{\prime\prime}_{2}$, $k^{\prime\prime}_{3}$ 
(for $\theta_{14}$ and $a_{4}$). Easy to prove that these
conditions are also sufficient for the periodicity of the 
resulting potential $V({\bf r})$.

 Theorem of Novikov permits to formulate here the following
property of the potentials $V_{B}^{eff}({\bf r})$ close enough 
to purely periodic potentials:

\vspace{0.5cm}

 {\bf Theorem 4.} {\it Consider a purely periodic potential
$V_{B}^{(0)eff}({\bf r})$ built by four interference pictures with 
the directions and periods $(\bm{\eta}^{(0)}_{1}, a^{(0)}_{1})$,
$(\bm{\eta}^{(0)}_{2}, a^{(0)}_{2})$, 
$(\bm{\eta}^{(0)}_{3}, a^{(0)}_{3})$,
$(\bm{\eta}^{(0)}_{4}, a^{(0)}_{4})$. Then there exists
such small region $\Gamma$ of parameters 
$\bm{\eta}_{1}$, $\bm{\eta}_{2}$, $\bm{\eta}_{3}$,
$\bm{\eta}_{4}$, $a_{1}$, $a_{2}$, $a_{3}$, $a_{4}$,
$I_{1}$, $I_{2}$, $I_{3}$, $I_{4}$ containing the initial potential
$V_{B}^{(0)eff}({\bf r})$ that for all the generic potentials 
$V_{B}^{eff}({\bf r})$ corresponding to the point of $\Gamma$ the 
following statements are true:

 1) All the non-singular open trajectories lie in the straight
strips of finite width and pass through them.

 2) All the regular trajectories have the mean direction in 
${\mathbb R}^{2}$ given by the equation

$$m_{1} X^{\prime} ({\bf r}) + m_{2} Y^{\prime} ({\bf r}) +
m_{3} Z^{\prime} ({\bf r}) + m_{4} W^{\prime} ({\bf r}) = 0 $$
with some integer (indivisible) $4$-tuple 
$(m_{1}, m_{2}, m_{3}, m_{4})$ which is the same for all the
(generic) points of ``stability zone'' $\Gamma$.

 3) The mean direction of open trajectories are the same for generic
potentials belonging to the same orbit of ``quasiperiodic group''. }

\vspace{0.5cm}

 Using Novikov theorem it's possible to prove also that the 
$4$-tuples $(m_{1}, m_{2}, m_{3}, m_{4})$ can be also defined
through the action of ``quasiperiodic group'' in the same way as in 
the case of 3 quasiperiods. 

 The asymptotic behavior of conductivity tensor $\sigma^{ik}$
is also the same in this case by the same reasons and the mean
directions of the open trajectories (and the integer $4$-tuples
$(m_{1}, m_{2}, m_{3}, m_{4})$) can be measured experimentally.

 According to Novikov theorem the regions with 
``topologically regular'' behavior can be found in any 
(arbitrarily small) open region of parameters 
$\bm{\eta}_{1}$, $\bm{\eta}_{2}$, $\bm{\eta}_{3}$,
$\bm{\eta}_{4}$, $a_{1}$, $a_{2}$, $a_{3}$, $a_{4}$,
$I_{1}$, $I_{2}$, $I_{3}$, $I_{4}$ and $B$. However, unlike the 
case $N = 3$ there is no theorem here restricting the existence 
of ``chaotic'' trajectories only to the case of just one energy 
level ($\epsilon_{1} = \epsilon_{2} = \epsilon_{0}$)
containing open trajectories. As we also mentioned already,
the case $N = 4$ is much more complicated from topological point
of view and there is no general classification of open 
trajectories in this case at the time. It is not clear also
if the topologically regular behavior
corresponds here to the generic situation or not and the
probability to find the chaotic behavior is unknown for 
this situation. 

 Let us now make some more general remark about the Novikov
problem in connection with 2D potentials $V({\bf r})$.
As can be seen, potentials $V({\bf r})$ with rather many
quasiperiods can be considered also as an interesting model
of random potentials on the plane. This model is rather
different from the standard models of random potentials
but still can have common features with them for big $N$
when the chaotic behavior of the open trajectories appears.
However, there is no strict theorems now which could connect
Novikov problem with the problems of random potentials on
the plane.

\section{Conclusion.} 

 We considered a special type of 
superlattices modulations giving quasiperiodic potentials
$V({\bf r})$ and $V_{B}^{eff}({\bf r})$
on the plane. For this type of potentials we 
considered the ``geometric limit'' ($\tau \rightarrow \infty$) of 
conductivity in the presence of magnetic field based on the
global geometry of the level curves of $V_{B}^{eff}({\bf r})$. 
The main attention was paid to the so-called ``topologically regular'' 
behavior of non-singular open level curves for the cases of
potentials with 3 and 4 quasiperiods. It was shown that it is 
possible to introduce ``topological numbers'' characterizing
the asymptotic behavior of $\sigma^{ik}$ similar to the numbers
introduced previously in the theory of normal metals. For the case
of 3 quasiperiods it was possible to give also the description
of structure of space of parameters giving potentials 
$V_{B}^{eff}({\bf r})$
according to the topological type of their non-singular open 
level curves. For the case of 4 quasiperiods only the part of the 
space of parameters corresponding to potentials close to
``purely rational'' was considered. It was shown that the 
corresponding ``topological numbers'' having the form of the
integer 4-tuples can be also introduced in this case.

 The author is grateful to Prof. S.P.Novikov for many fruitful
discussions on this problem. The author is also grateful
to Prof. I.A. Larkin who brought the articles on 2D electron 
gas to his attention for the interest to this work and advice.

\end{document}